\documentclass[fleqn,usenatbib]{mnras}
\usepackage{amsmath}
\usepackage{newtxtext,newtxmath}
\usepackage[T1]{fontenc}
\usepackage{ae,aecompl}

\usepackage{graphicx}	% Including figure files

\title[Eccentricity and inclination of hot protoplanets]{Evolution of eccentricity and inclination of hot protoplanets embedded in radiative discs}

\author[H. Eklund $\&$ F. Masset]{
Henrik Eklund,$^{1}$\thanks{E-mail: eklhen@gmail.com}
Fr\'ed\'eric S. Masset$^{2}$
\\
$^{1}$Department of Physics, Gothenburg University, Origov\"agen 6 b, 41296 Gothenburg, Sweden\\
$^{1, 2}$Instituto de Ciencias F\'isicas, Universidad Nacional Aut\'onoma de M\'exico, Av. Universidad s/n, 62210 Cuernavaca, Mor., Mexico\\
}

\date{Accepted 2017 April 4. Received 2017 April 2; in original form 2017 January 30}

\pubyear{2017}

\begin{document}
\label{firstpage}
\pagerange{\pageref{firstpage}--\pageref{lastpage}}
\maketitle

\begin{abstract}
  We study the evolution of the eccentricity and inclination of
  protoplanetary embryos and low-mass protoplanets (from a fraction of
  an Earth mass to a few Earth masses) embedded in a protoplanetary
  disc, by means of three dimensional hydrodynamics calculations with
  radiative transfer in the diffusion limit. When the protoplanets
  radiate in the surrounding disc the energy released by the accretion
  of solids, their eccentricity and inclination experience a growth
  toward values which depend on the luminosity to mass ratio of the
  planet, which are comparable to the disc's aspect ratio and which
  are reached over timescales of a few thousand years.  This growth is
  triggered by the appearance of a hot, under-dense region in the
  vicinity of the planet.  The growth rate of the eccentricity is
  typically three times larger than that of the inclination. In long
  term calculations, we find that the excitation of eccentricity and
  the excitation of inclination are not independent. In the particular
  case in which a planet has initially a very small eccentricity and
  inclination, the eccentricity largely overruns the inclination. When
  the eccentricity reaches its asymptotic value, the growth of
  inclination is quenched, yielding an eccentric orbit with a very low
  inclination.

  As a side result, we find that the eccentricity and inclination of
  non-luminous planets are damped more vigorously in radiative discs
  than in isothermal discs.
\end{abstract}


%%%%%%%%%%%%%%%%% BODY OF PAPER %%%%%%%%%%%%%%%%%%

\section{Introduction}

When a low-mass protoplanet perturbs its parent protoplanetary disc
solely by gravity, its eccentricity and inclination are usually
damped. The damping time scale of these quantities is, for an Earth
mass planet embedded in a disc similar to the Minimum Mass Solar
Nebula, of order of a thousand years at a few astronomical units. This
small time scale is shorter, by a factor of $h^{-2}$ ($h$ being the
disc's aspect ratio) than the migration time scale \citep{arty93b}.
In the limit of vanishing eccentricity and inclination, the damping
rates can be estimated by an expansion of the perturbing potential to
first order in $e$ and $i$. This was the approach of
\citet{2004ApJ...602..388T} who evaluated these rates for a planet in
an isothermal disc by means of semi-analytical, linear
calculations. When these orbital elements are not negligible compared
to the disc's aspect ratio, it is necessary to include higher order
terms. This was done by \citet{paplar2000}, who included all
resonances necessary for convergence, and who considered for the first
time eccentricities larger than the disc's aspect ratio. For very high
eccentricities, however, a resonant approach becomes unpractical and
one may rather resort to a dynamical friction calculation
\citep{pap2002, 2011ApJ...737...37M}, as the disc response is local,
and the Keplerian shear unimportant.  Numerical simulations have been
employed to confirm and extend analytical predictions
\citep{2007A&A...473..329C}, and have allowed to relax the customary
isothermal approximation \citep{2010A&A...523A..30B,
  2011A&A...530A..41B}. In all cases the disc's tide has been found to
damp eccentricity and inclination.

Not all the disturbances imparted to the disc by a protoplanetary
embryo are due to gravity. \citet{2015Natur.520...63B} have recently
investigated the impact on planetary migration of heat release by a
hot, accreting embryo in the surrounding disc, and found that this
effect is able revert the migration of embryos up to a few Earth
masses. This study was however limited to planets on circular orbits.
By a dynamical friction calculation, \citet{2017MNRAS.465.3175M} have
studied a perturber that travels across a uniform, gaseous and opaque
medium, and that releases heat at a fixed rate in the surrounding
gas. They found that the hot, underdense trail thus triggered in the
gas exerts a force on the perturber (dubbed the heating force)
directed along its motion, and independent of the perturber's velocity
in the limit of low Mach numbers. Applying this result to embedded
protoplanetary embryos, they suggest these could reach eccentricities
and inclinations comparable to the disc's aspect ratio.  Whether a
dynamical friction calculation in a disc is justified, however,
depends on the competition between two time scales: the response time
scale and the shear time scale. When the latter is larger than the
former, the shear is unimportant and one can resort to a calculation
of dynamical friction. In the case of the hot plume, this should occur
for values of the eccentricity largely smaller than the disc's aspect
ratio \citep{2017MNRAS.465.3175M}. We will confirm this expectation in
the present work. The disc's tide\footnote{Throughout this paper, we
  call disc's tide the force exerted on the planet by the perturbation
  that would be induced in the disc if the planet was non-luminous.},
however, can only be captured by a dynamical friction calculation for
eccentricities in excess of the disc's aspect ratio
\citep{2002A&A...388..615P,2011ApJ...737...37M}. The asymptotic values
of eccentricity and inclination quoted by \citet{2017MNRAS.465.3175M},
obtained through a dynamical friction calculation both for the hot
trail and for the tide of the ambient gas, may therefore not be
accurate, as they are comparable to the disc's aspect ratio.  The
purpose of the present work is to investigate, by means of numerical
simulations, whether hot, embedded embryos are indeed subjected to an
eccentricity and inclination growth, and which values these orbital
elements can reach at larger time.

Our paper is organised as follows. In Section~\ref{sec:Disc} we
describe our setup and the numerical methods employed. In
Section~\ref{sec:prel study} we investigate the eccentricity and
inclination behaviour of non-luminous embryos, firstly in isothermal
discs so as to validate our numerical procedures by comparing our
code's outcome to the analytical expressions of
\citet{2004ApJ...602..388T}, and secondly in a radiative disc similar
to the one that will be considered throughout the rest of the paper.
In Section~\ref{sec:results} we present our results for luminous
embryos. These consist of fiducial calculations (one for the
eccentricity, one for the inclination), and of different explorations
of parameter space. Namely we explore the impact of the initial value
of eccentricity or inclination, the planetary mass, and the mass
accretion rate. We also present long term calculations, spanning
$2000$~or $2500$~orbital periods, in which we consider planets that
are simultaneously eccentric and inclined. We discuss our results in
Section~\ref{sec:discussion} and draw our conclusions in
Section~\ref{sec:conclusions}.
\section{Setup}
\label{sec:Disc}
In this section we present the different parts of our setup and the
numerical methods employed.
\subsection{Disc}
\label{sec:disc} % used for referring to this section from elsewhere
We consider a protoplanetary disc around a central star of mass
$M_\star$. We denote its aspect ratio with $h=H/r$, where $H$ is the
pressure scale length of the disc and $r$ is the radial distance from
the star. The surface density of the disc is given by:
\begin{equation}
\label{eq:1}
\Sigma(r)=\Sigma_0 \left(\frac{r}{r_0}\right)^{-p},
\end{equation} 
where $\Sigma_0$ is the surface density at $r=r_0$ and where we adopt
a slope of surface density $p=1$. The disc has a constant opacity
$\kappa=1.0$ cm$^2$ g$^{-1}$ and a constant kinematic viscosity
$\nu=1.016\times10^{15}$ cm$^2$ s$^{-1}$, which translates into an
$\alpha$-value \citep{ss73} of $4\cdot 10^{-3}$ for our isothermal
discs, and $5.6\cdot 10^{-3}$ for our radiative discs, which have a
slightly different thickness. In this first, exploratory work, the
opacity is kept to this fixed value in order not to introduce an
additional, complex dependence of the disc's thermal diffusivity on
its temperature and density.

\subsection{Planet}
\label{sec:planet}

A planet with mass $M_\mathrm{p}$ is inserted in the disc on an orbit
around the central star with a semi-major axis $a=5.2$~au
(astronomical units). We denote by $e$ and $i$ its eccentricity and
inclination, respectively, and by $e_0$ and $i_0$ the initial values
of these orbital elements. All values quoted for the inclination
throughout this paper are in radians. The planet is freely moving in
the disc. It is heated up by the bombardment of infalling objects,
such as planetesimals or pebbles. We assume the latent heat for
vaporisation of the infalling bodies to be a small fraction of their
potential energy, which is a reasonable assumption for planets in the
Earth mass range \citep{2015Natur.520...63B}. We also assume that if
infalling bodies are destroyed far above the planetary surface, their
debris eventually reach the surface, thereby ultimately releasing the
potential energy $GM_\mathrm{p}/R_\mathrm{p}$ per unit mass, where
$R_\mathrm{p}$ is the physical radius of the planet. This assumption
is known as the sinking hypothesis \citep{1996Icar..124...62P}.  The
luminosity of the planet thus reads:
\begin{equation}
  \label{eq:2} 
L=\frac{GM_\mathrm{p}\dot M_\mathrm{p}}{R_\mathrm{p}},
\end{equation}
where $G$ is the gravitational constant and $\dot M_\mathrm{p}$ the planetary accretion
rate. We introduce the mass doubling time:
\begin{equation}
  \label{eq:3}
\tau=\frac{M_\mathrm{p}}{\dot M_\mathrm{p}} ,
\end{equation}
so that the luminosity is given by:
\begin{equation}
  \label{eq:4}
  L=\frac{GM_\mathrm{p}^2}{R_\mathrm{p}\tau}.
\end{equation}
The planet's physical radius is evaluated assuming a homogeneous mass
distribution with density $\rho_\mathrm{p}=3$ $\text{g}$ $\text{cm}^{-3}$. Our results do
not depend sensitively on this assumption, since the planet's physical radius
scales with the cubic root of the mean density.

\subsection{Numerical setup}
\label{sec:numerical setup}
We use the public code
FARGO3D\footnote{\texttt{http://fargo.in2p3.fr}}
\citep{2016ApJS..223...11B} with orbital advection enabled
\citep{fargo2000}. The version of the code that we used includes a
non-public radiative transfer module.  We use a spherical mesh with
azimuthal extent $[-\phi_\mathrm{max},\phi_\mathrm{max}]$, where
$\phi_\mathrm{max}$ depends on the run and will be specified
later. The radial extent is $[3/5a, 7/5a]$ and the extent in
colatitude is $[\pi/2-0.12, \pi/2+0.12]$. We use $N_r=512$ cells in
radius, $N_\theta=128$ cells in colatitude, and a number of cells in
azimuth $N_\phi$ that depends on $\phi_\mathrm{max}$: when
$\phi_\emph{max}=\pi$ (i.e. when we simulate a full disc), we use
$N_\phi=1024$~cells. Our resolution respectively in azimuth, radius
and colatitude is therefore:
$(\Delta \phi,\Delta r,\Delta\theta)=(6.1\cdot 10^{-3},1.6\cdot
10^{-3}a, 1.9\cdot 10^{-3})$. We soften the planetary potential over a
length $\epsilon=1.5\cdot 10^{-3}a$, comparable to our radial
resolution. When $\phi_\mathrm{max}$ differs from $\pi$, we adjust the
number of cells in azimuth so as to keep $\Delta\phi$ constant. When
simulating radiative discs, radiative transfer is dealt with in the
same manner as in \citet{2015Natur.520...63B}. We use
wavelength-independent flux limited diffusion with a two-temperature
approach \citep{2013A&A...549A.124B}. The frame in which our
calculations are performed corotates with the planet, hence it has a
time-varying rotation rate whenever the planet is either eccentric or
inclined.

At each timestep $\Delta t$, the energy released by the planet
$\Delta E=L\Delta t$ is used to increase the internal energy in the
cells surrounding the planet, in a separate substep. The distribution
of $\Delta E$ among each of the eight neighbouring cells is determined
using a virtual cell centred on the planet, with same size as the
neighbouring cells. The fraction of this cell's volume occupied by
each of the neighbours determines the fraction of $\Delta E$ that is
attributed to this neighbour. This procedure implies that the
barycentre of the energy released lies at the planet location. In the
calculation of the different fractions, we neglect the mesh curvature
since the energy release is a very local process that takes place on
eight neighbouring cells at most. Our heat release occurs on the
smallest possible scale allowed by the mesh. The force exerted by the
heated region will be accurately captured if the size of this region
is much larger than the resolution of the mesh. We will study this
requirement in more detail in section~\ref{sec:temp-pert-disc}.

\section{Preliminary study}
\label{sec:prel study}
In a first step, we check the behaviour of the planet without heat
release. We do this firstly in a locally isothermal disc, so as to
compare our results with the analytic estimates of
\citet{2004ApJ...602..388T}, then in a radiative disc.  In each case
six runs are performed in which the orbit of the planet is initially
set either with eccentricities $e=0.01$, $0.015$ and $0.02$ or with
inclinations $i=0.01$, $0.015$ and $0.02$ radians. The planet is free
to move in all direction. However we spawn eccentric planets 
coplanar with the disc, and inclined planets on circular
orbits, so as to study the time evolution of the orbital elements in a
separate manner. Our simulations are evolved for 5 orbits of the
planet only. This amount of time appears to be sufficient to allow a
precise measurement of the variation rate of the orbital elements
under study.

\subsection{Isothermal disc}
\label{sec:isothermal-disc}
In this study we use a disc with a uniform aspect ratio $h=0.05$, while our
planet has a mass $3\cdot 10^{-6}M_\star$, which translates into one Earth mass if
the central star has a solar mass.
We cast our results in terms of the characteristic time \citep{2004ApJ...602..388T}:
\begin{equation}
\label{eq:5}
t_c =q^{-1}\left( \frac{\Sigma_0 a^2}{M_\star}\right)^{-1} \left( \frac{c_s}{a \Omega_\mathrm{p}} \right)^4 \Omega_\mathrm{p}^{-1},
\end{equation} 
where $q=M_\mathrm{p}/M_\star$ and $c_s$ is the sound speed at the radius of the planet and $\Omega_\mathrm{p}$ the
planet's orbital frequency. 
\begin{figure*}
  \centering
  \includegraphics[width=.24\textwidth]{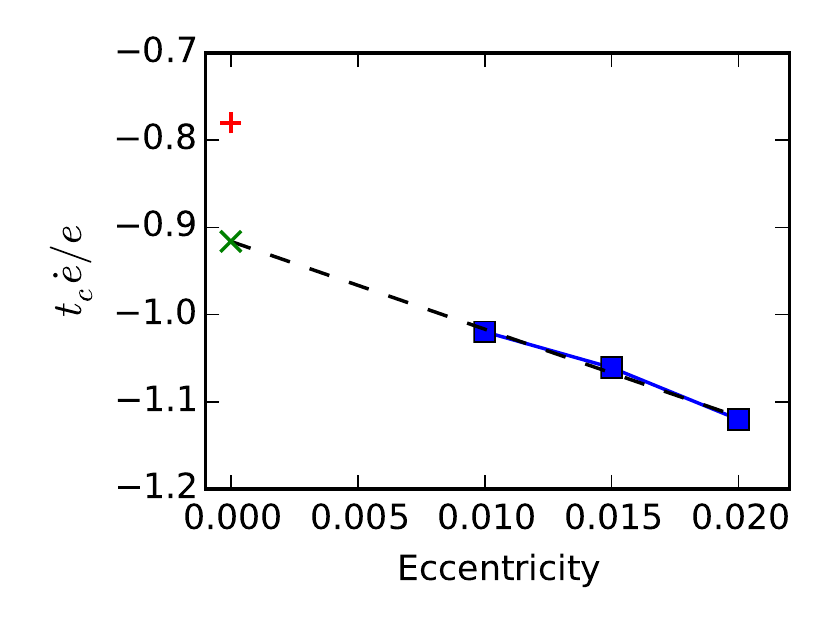}
  \includegraphics[width=.24\textwidth]{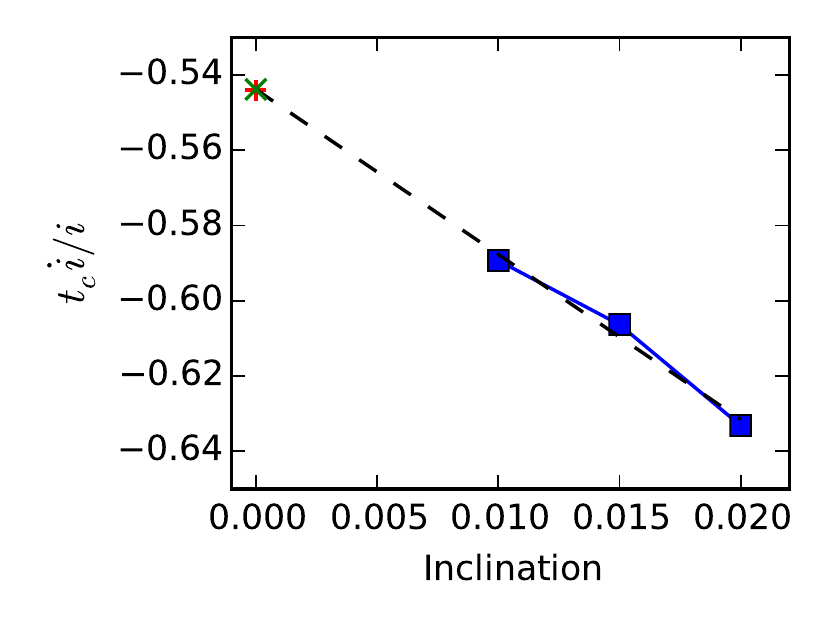}
  \includegraphics[width=.24\textwidth]{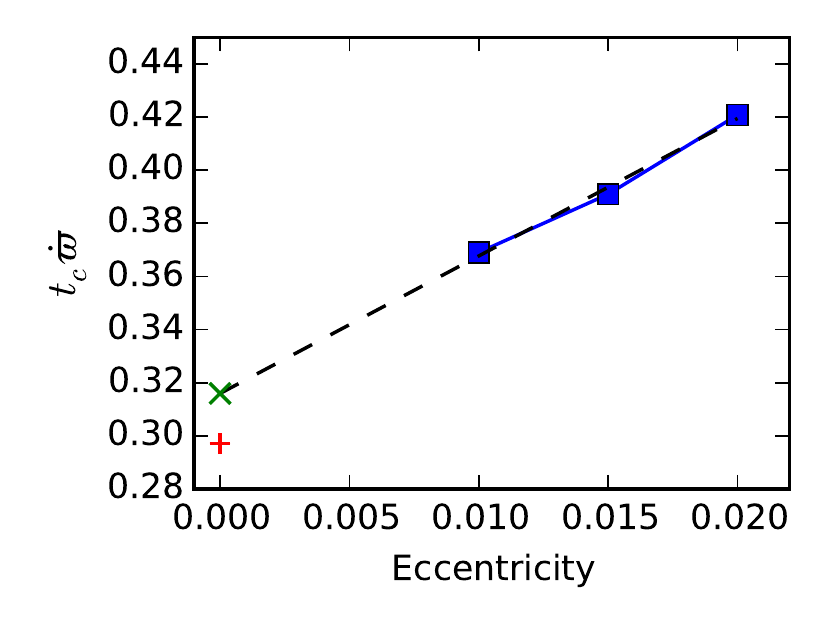}
  \includegraphics[width=.24\textwidth]{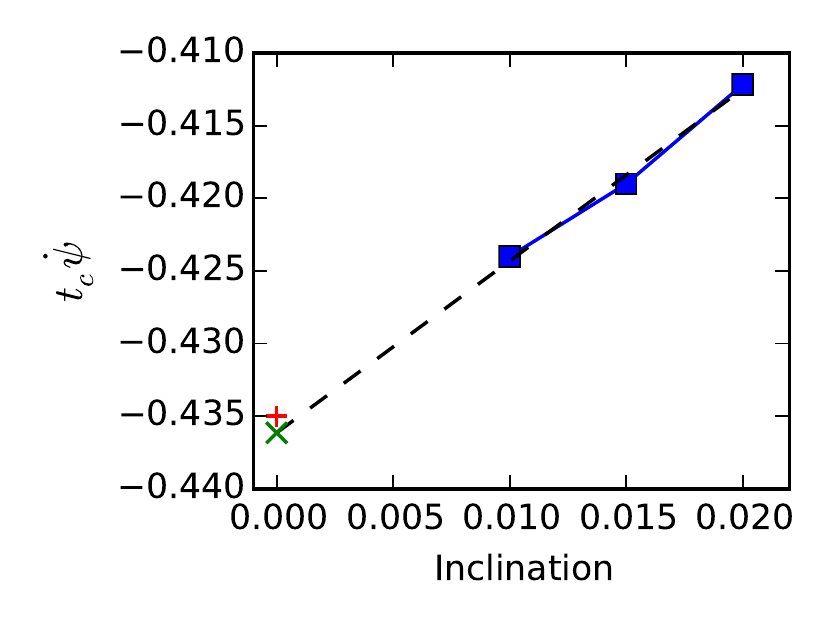}
  \caption{\label{fig:tw04}From left to right: eccentricity damping
    rate, inclination damping rate, rate of precession of periapsis
    for an eccentric planet, and rate of precession of the line of
    nodes for an inclined planet, in an isothermal disc.  In each plot
    these quantities are displayed as a function either of
    eccentricity or inclination. Also shown with a $\times$ sign is
    the value for a vanishing eccentricity or inclination,
    extrapolated from the simulations data with a linear regression
    fit (dashed lines). We show with a $+$ sign the value given by
    \citet{2004ApJ...602..388T}.}
\end{figure*}
Our results are shown in Fig.~\ref{fig:tw04}. We also determine the
precession rates of the periapsis and of the line of nodes,
respectively for eccentric and inclined cases. Our largest difference
with the results of \citet{2004ApJ...602..388T} is found on the
eccentricity damping rate, and amounts to $17$~\%. The vertical extent
of our mesh, which is $\pm 2.4h$, only contains $98$~\% of the column
density of the full disc, so we do not expect a match better than to a
few percents. Globally, these results are in very satisfactory
agreement with the analytic expressions of
\citet{2004ApJ...602..388T}.

\subsection{Radiative disc}
\label{sec:radiative-disc}
The study of heat release that will be presented in section~\ref{sec:results} is
performed in a radiative disc. Here we study the inclination and eccentricity
evolution of a planet without heat release in a radiative disc. Prior to the
simulations with a planet, the radiative disc is relaxed toward hydrostatic and
radiative equilibria by means of two-dimensional runs in the $(r,\theta)$ plane.
From now on we specialise to the case of a central object of solar mass
($M_\star=M_\odot$), and to the values $a=r_0=5.2$~au, $\Sigma_0=200$~g~cm$^{-2}$.
The ratio of specific heats at constant pressure and constant volume is
$\gamma=1.4$. After
convergence toward equilibrium, we obtain an aspect ratio at the planet location
of $h_r=0.042$, close to the value given by \citet{2013A&A...549A.124B}:
\begin{equation}
  \label{eq:6}
  h=\left(\frac{9}{32}\frac{{\cal R}^4}{\mu^4}\kappa\nu\Sigma_0^2\frac{r_0}{G^3M_\star^3\sigma}\right)^{1/8}=0.043
\end{equation}
where ${\cal R}$ the ideal gas constant, $\sigma$ the Stefan constant
and $\mu=2.3$~g~mol$^{-1}$ is the mean molecular weight. As the disc
mostly settles vertically its surface density profile is essentially
unchanged with respect to the initial conditions, given by
Eq.~\eqref{eq:1}. Once the disc has reached equilibrium, a planet of
one Earth mass ($M_\mathrm{p}=M_\oplus$) is inserted at $r_0=5.2$ au and let
free to evolve through the disc. Our setup is essentially similar to
that of \citet{2015Natur.520...63B}, except that we relax the
assumption of circular and coplanar orbits. As previously, the
variation of its orbital elements is subsequently measured over five
orbits. The rates of change measured for $e$ and $i$ differ much from
the isothermal case, as can be seen in column~3 of Table~\ref{tab:ecc
  and inc reduced disc}.  We normalise the damping rates with:
\begin{equation}
  \label{eq:7}
  t_c=q^{-1}\left(\frac{\Sigma_0a^2}{M_\star}\right)^{-1}h_r^4\Omega_\mathrm{p}^{-1}=3.2\mbox{~kyrs}
\end{equation}
The damping rates, both for the eccentricity and inclination, are
significantly in excess of those found in isothermal discs (see
section \ref{sec:isothermal-disc}), and they display a trend to be
larger when the corresponding orbital element is smaller. This is in
sharp contrast with the results of \citet{2010A&A...523A..30B},
although these are not directly comparable to ours, since these
authors considered a large mass planet with $q/h^3\sim 1.2$, whereas
we are concerned here with low mass, embedded objects. This behaviour is the subject of current investigation, the results of which will be presented in a forthcoming
publication. We note that
$h_r=c_s^\mathrm{iso}/(r\Omega_\mathrm{p})$, where $c_s^\mathrm{iso}$ is the
isothermal sound speed, so that Eq.~\eqref{eq:7} is equivalent to
Eq.~\eqref{eq:5} when $c_s\equiv c_s^\mathrm{iso}$. If we had used the
adiabatic sound speed in Eq.~\eqref{eq:7}, the time scale $t_c$ would
be larger by a factor $\gamma^2\sim 2$. This would compound the
difference with the isothermal results. We stress that our choice of
the expression of $t_c$ is only a choice of normalisation which does
neither affect our conclusions nor the outcomes of the runs.

\subsubsection{Reduction of the azimuthal extent}
\label{sec:reduct-azim-extent}
Since our study involves a significant exploration of parameter space,
the reduction of the cost per simulation is a concern. We investigate
how the results for a full disc in azimuth ($\phi_\mathrm{max}=\pi$)
are affected when one considers only a quadrant in azimuth
($\phi_\mathrm{max}=\pi/4$), keeping the same resolution (using
therefore only 256 cells in this direction instead of 1024). When
doing so it is important that the frame corotates with the planet (so
that the planet lies on the bisector of the quadrant at any instant in
time), and that the azimuthal averaged density be subtracted from each
cell prior to the evaluation of the disc's force onto the planet
\citep{2008ApJ...678..483B}.

Calculations restricted to a quadrant give the evolution rates of
Table~\ref{tab:ecc and inc reduced disc} in column~4. The eccentricity
damping rates display changes with respect to the full disc
calculations at the percent level, while the inclination damping rates
are reduced by approximately $10$~\%.  We consider these variations
acceptable, and perform the rest of our simulations on a quadrant.

\begin{table} \centering \caption{Eccentricity and inclination damping
    rates in radiative discs simulated over a full mesh (F) and over a
    quadrant only (Q).  The second column shows the analytic estimates
    of \citet{2004ApJ...602..388T} for low values of the eccentricity
    and inclination.  The rates measured in the simulations are given
    for three different initial values of the corresponding orbital
    element. Column~3 gives the evolution rates for the full disc,
    column~4 for the quadrant and column~5 shows the ratio of
    columns~3 and~4.}
	\label{tab:ecc and inc reduced disc}
	\begin{tabular}{ccccc} % five columns, alignment for each 
		\hline 
$e_0$ & $t_c\left.\frac{\bar{\dot{e}}}{e}\right|_{TW04}$ &$t_c\left.\frac{\bar{\dot{e}}}{e}\right|_F$ & $t_c\left.\frac{\bar{\dot{e}}}{e}\right|_Q$ & $\frac{{(\bar{\dot{e}}/e)}_F}{(\bar{\dot{e}}/e)_Q}$ \\
\hline  

$\rightarrow 0$ & -0.78 &  & &\\
0.010 && -1.75 & -1.73 & 1.01\\

0.015 && -1.45 & -1.44 & 1.01\\

0.020 && -1.27 & -1.26 & 1.01\\
		\hline 
		\hline 
$i_0$ & $t_c\left.\frac{\bar{\dot{i}}}{i}\right|_{TW04}$ &$t_c\left.\frac{\bar{\dot{i}}}{i}\right|_F$ & $t_c\left.\frac{\bar{\dot{i}}}{i}\right|_Q$ & $\frac{{(\bar{\dot{i}}/i)}_F}{(\bar{\dot{i}}/i)_Q}$ \\
\hline 
$\rightarrow 0$ & -0.54 &  & &\\ 
0.010 && -0.97 & -0.89 & 1.09\\ 
0.015 && -0.85 & -0.77 & 1.10\\
0.020 && -0.77 & -0.69 & 1.11\\
\hline 
	\end{tabular}
\end{table}

We comment that the impact of the circular resonances\footnote{The
  resonances that exist even when the planet is on a circular orbit, such as the
  coorbital corotation resonances or the Lindblad resonances
  associated with the angular frequency of the planet's guiding
  centre. These are the resonances that drive the migration of a
  planet on a nearly circular orbit.} on the eccentricity evolution is
marginal. The driving or damping timescales of these resonances are
comparable to the migration timescale, hence a factor $h^{-2}$ larger
than the timescale associated to the eccentric Lindblad and corotation
resonances, which are first order in the eccentricity, and which play
the most important role for a non-luminous planet. We therefore
anticipate that an accurate description of the torque arising from the
circular resonances should be unimportant for the process we describe
in this work, which occurs on time scales shorter than that of
migration. This comment applies in particular to the coorbital
corotation torque or its non-linear version, the horseshoe drag
\citep{2009arXiv0901.2265P}. Nonetheless, it should be kept in mind
that studies of the migration of hot planets with our setup (which is
not the primary scope of this work) may be biased by two effects:
\begin{itemize}
\item Working on a restricted domain in azimuth shortens the horseshoe
  libration time, and consequently affects the ratio of this time to
  the viscous time across the horseshoe region, which controls the
  degree of saturation of the horseshoe drag
  \citep{masset01,2010ApJ...723.1393M,pbk11}.
\item The horseshoe region is not resolved for planets with masses
  lower than $O(10^{-1})\;M_\oplus$, resulting in an incorrect tidal
  torque from the disc. For such low masses, however, migration is
  very slow and hardly relevant at all.
\end{itemize}

\section{Study of heat release}
\label{sec:results}
We present in this section the evolution rates of
eccentricity and inclination of a planet that is releasing heat into
the surrounding disc. In section~\ref{sec:fiducial calculation} we
firstly present our fiducial calculations. Subsequently, we explore the
behaviour of the evolution rates as a function of several
parameters. These parameters are the initial values of the
eccentricity and the inclination (section~\ref{sec:varying initial
  values}), the planetary accretion rate (section~\ref{sec:mass
  doubling time}) and the planetary mass (section~\ref{sec:planetary
  mass}). All the evolution rates presented here are mean values over
the first five orbits after the planet is inserted into the disc. In
section~\ref{sec:temp-pert-disc} we examine the temperature excess due
the heat release for an eccentric and an inclined planet. Finally, in
section~\ref{sec:long term behavior} we explore the long-term
behaviour of the eccentricity and inclination of a hot planet.

\subsection{Fiducial calculations}
\label{sec:fiducial calculation}
Our two fiducial calculations (one for the eccentricity, another one
for the inclination) have same parameters as the ones used in
section~\ref{sec:radiative-disc}. In addition, we set for the planet a
mass doubling time of $10^5$~yrs, which determines its luminosity
through Eq.~\eqref{eq:4}. This mass doubling time lies within the
admitted range of mass doubling times for Earth-sized protoplanets at a
distance of $5.2$~au, and represents a conservative value typical of
planetesimal accretion \citep{1996Icar..124...62P} rather than the
more efficient pebble accretion \citep{2012A&A...544A..32L}. The full
set of parameters is summed up in Table~\ref{tab:fidpar} for the
reader's convenience.
%Whenever we study the dependence of  the evolution rates as a function of a given parameter : say that later.

\begin{figure}
  \includegraphics[width=\columnwidth]{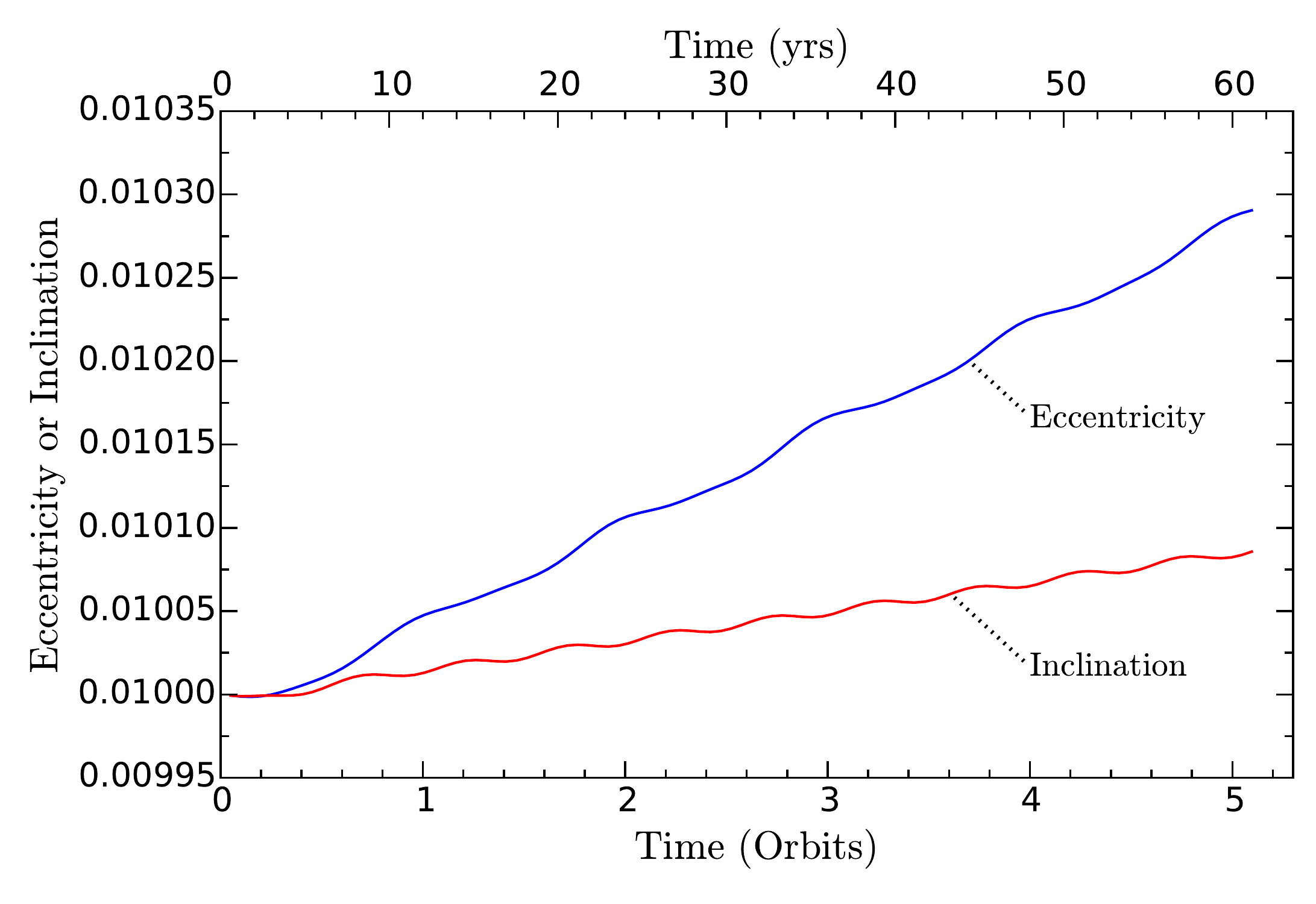}
  \caption{Evolution of the eccentricity and the inclination of the luminous planet of the fiducial calculations.}
  \label{fig:fiducial run}
\end{figure}

\begin{table*} \centering \caption{Parameters of the fiducial calculations. The
    planet is inserted in the disc and released at apocentre or on its ascending
    line of nodes once the disc has reached hydrostatic and thermal equilibrium,
    at which point the disc has an aspect ratio $h_r=0.042$ at the location of
    the planet.}
	\label{tab:fidpar}
	\begin{tabular}{|rlrlrl|} % four columns, alignment for each 
		\hline 
Parameter & Value & Parameter & Value & Parameter & Value\\
\hline 
$\mu$ & $2.3$~g~mol$^{-1}$ &$p$ & $1$&$N_\phi$& 256 cells\\
$\rho_\mathrm{p}$ & $3$~g~cm$^{-3}$& $\tau$ & $10^5$~yrs &$\phi$ interval &$[-\pi/4,\pi/4]$\\
$\Sigma_0$ & $200$~g~cm$^{-2}$ & $M_\star$ & $1\;M_\odot$ &$N_r$& 512 cells\\
$(e_0,i_0)$ &  $(0.01,0)$ or $(0,0.01)$ &$r_0$, $a$ & $5.2$~au &$r$ interval (au)& $[3.12, 7.28]$ \\
$\kappa$ & $1$~cm$^2$~g$^{-1}$ &$M_\mathrm{p}$ & $1\;M_\oplus$ & $N_\theta$& 128 cells\\
$\nu$ & $1.016\cdot 10^{15}$~cm$^2$~s$^{-1}$ &$\gamma$ & $1.4$ &$\theta$ interval &$[\pi/2-0.12,\pi/2+0.12]$\\
		\hline 
	\end{tabular}
\end{table*}
The time evolution of the eccentricity and inclination is shown in
Figure $\ref{fig:fiducial run}$. Both quantities steadily
increase. The rate of change of the eccentricity is
$t_c\bar{\dot{e}}/e = +1.57$, that of the inclination is
$t_c\bar{\dot{i}}/i = +0.48$ and thus the eccentricity broadly
increases three times faster than the inclination.

\subsection{Varying the initial value of eccentricity or inclination}
\label{sec:varying initial values}
We study hereafter the dependence of the driving rate of either the
eccentricity or inclination upon its initial value. An immediate
application of such study is that it allows to mimic, in a piecewise
fashion, a long-term evolution by connecting one after another short
term graphs extrapolated to the next available initial eccentricity or
inclination, so as to infer the behaviour of the orbital element at
larger time (this implicitly assumes that the driving rate of the
orbital element depends only on its instantaneous value, rather than
on its history).

\begin{figure}
	\includegraphics[width=\columnwidth]{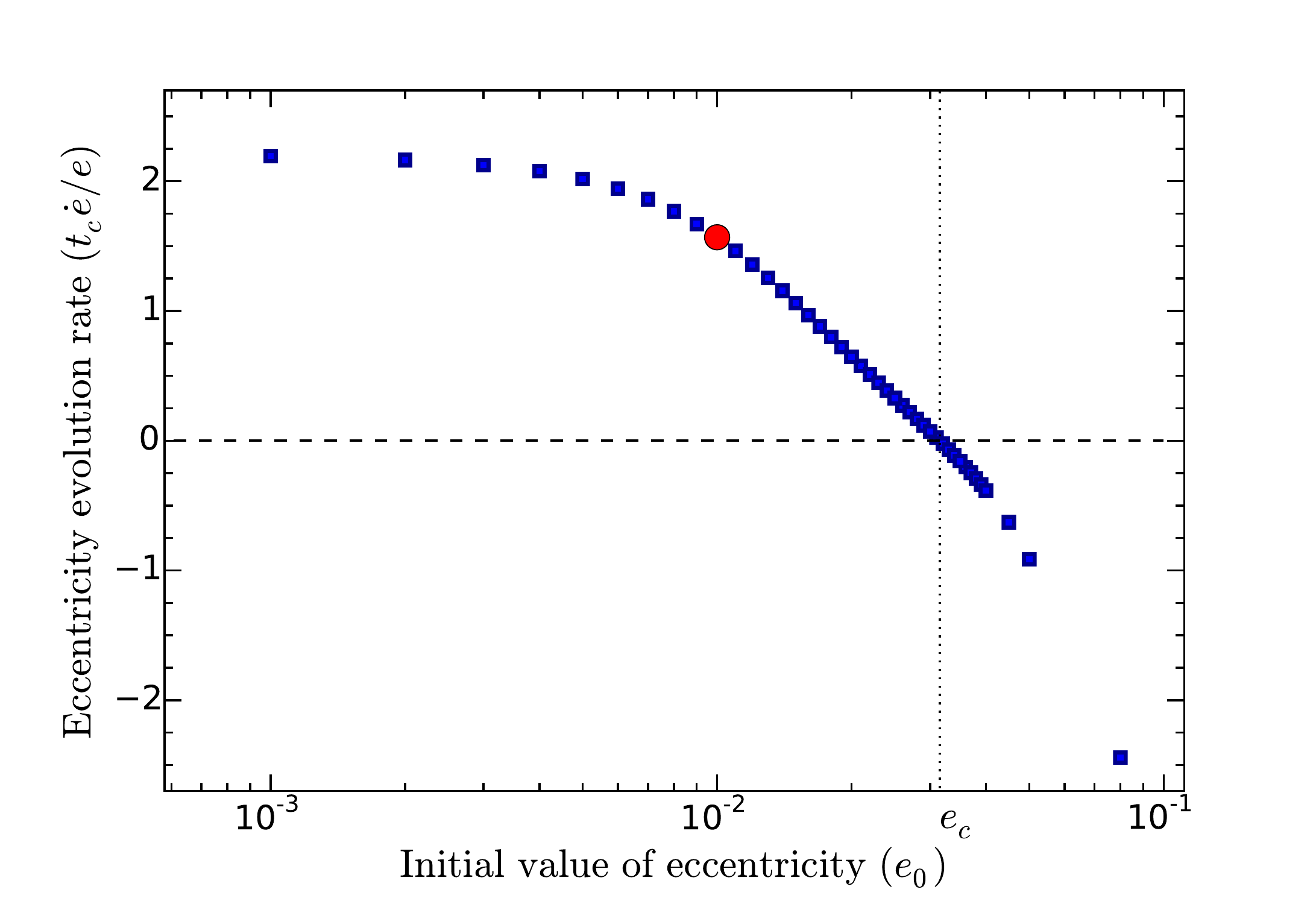}
        \caption{Time derivative of eccentricity as a function of eccentricity. The
          fiducial calculation is shown with a red disc. The eccentricity is
          found to have a null time derivative for the critical value
          $e_c\approx 0.0315$.}
    \label{fig:ecc variation}
\end{figure}

\begin{figure}
	\includegraphics[width=\columnwidth]{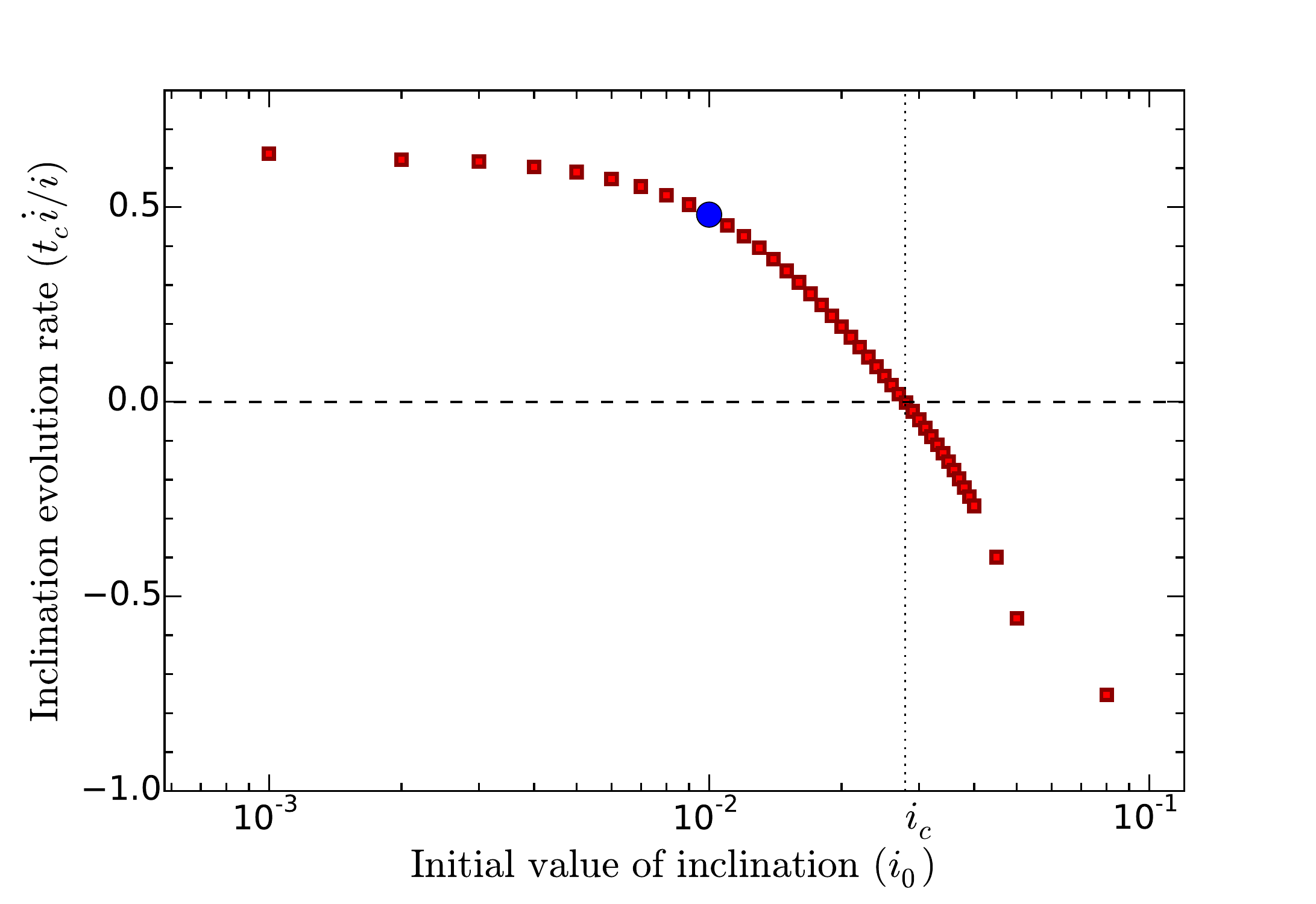}
        \caption{Time derivative of inclination as a function of inclination.
          The fiducial calculation is shown with a blue disc. The inclination is
          found to be constant in time at the critical value
          $i_c\approx 0.028$.}
    \label{fig:inc variation}
\end{figure}

We vary the eccentricity and inclination with respect to their
fiducial values as follows. They are set to values ranging between
$0.001$ and $0.04$ by steps of $0.001$ and to $0.045$, $0.05$ and
$0.08$. The results are presented in Figs.~\ref{fig:ecc variation}
and~\ref{fig:inc variation}.

We recover the fact that the eccentricity time derivative is broadly
three times larger than that of the inclination. Both graphs show an
horizontal asymptote at low values ($\lesssim 0.01$), which suggests
that both the eccentricity and inclination undergo an exponential
growth up to $\sim 0.01$. The eccentricity time derivative shows
a monotonous behaviour, and a sign reversal at $e=e_c$. This suggests
that the eccentricity, if initially smaller than $e_c$, will grow and
tend toward $e_c$ at larger time. Reciprocally, if it is initially
larger than $e_c$, it should decay toward $e_c$ at larger time. Since
$t_c\dot e/e=O(1)$, one can infer that the characteristic timescale
for the variation of $e$ is $t_c$, which amounts here to a few
thousand years, as evaluated in Eq.~\eqref{eq:7}. Similar
considerations apply to the inclination, which tends toward a value
marginally smaller than $e_c$.

We note in Fig.~\ref{fig:inc variation} that the decay rate of the
most inclined case is smaller than could be extrapolated from the data
at lower inclination. This is presumably due to the fact that for this
case, the planet is out of the disc for a significant fraction of its
orbit.  This behaviour, in contrast, is not expected, and not found,
for the eccentricity (Fig.~\ref{fig:ecc variation}).

\subsection{Dependence on the accretion rate}
\label{sec:mass doubling time}
\begin{figure}
  \includegraphics[width=\columnwidth]{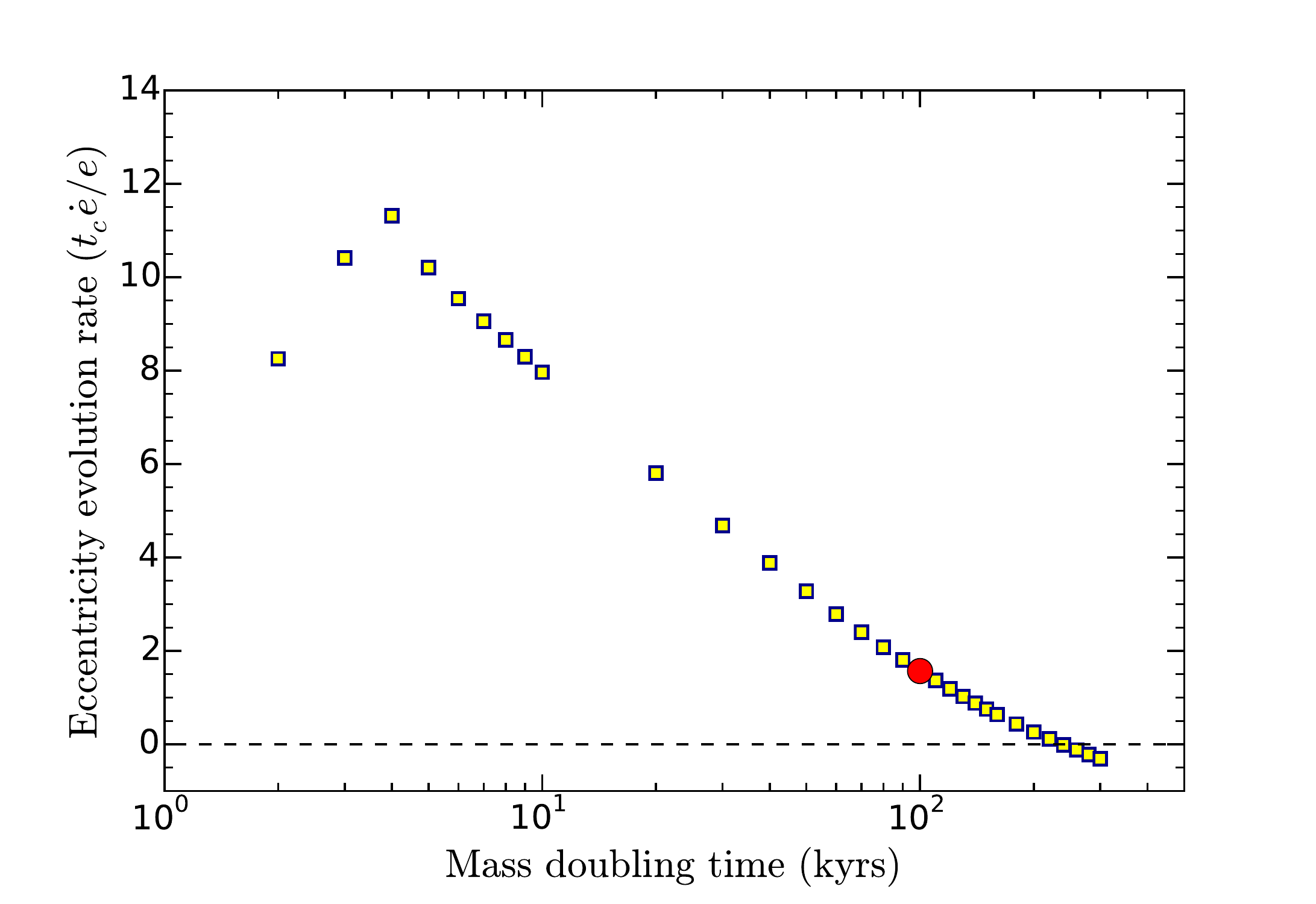}
  \caption{Growth or damping rate of eccentricity as a function of the mass
    doubling time. The fiducial calculation is indicated with a red
    disc. For all calculations $e_0=0.01$.}
   \label{fig:ecc accretion rate}
\end{figure}

The mass doubling time $\tau$ is varied between $1$~kyrs and
$300$~kyrs. The orbital parameters change over a timescale much
shorter than the mass doubling time. Thus we do not add the accreted
mass to the planetary mass.  For each case the mean evolution rates of
the eccentricity and inclination are calculated. The eccentricity
evolution rates are presented in Figure \ref{fig:ecc accretion rate}
and the inclination evolution rates in Figure \ref{fig:inc accretion
  rate}. With this specific planet-disc setup of the fiducial run, the
eccentricity increases for mass doubling times shorter than about
$240$~kyrs. The inclination needs a slightly larger luminosity and
increases for mass doubling times shorter than about $200$~kyrs. With
longer mass doubling times than these, i.e. slower mass accretion or
lower luminosities, the luminosity is not large enough to yield an
eccentricity or inclination growth, and these orbital parameters would
decay below their fiducial initial value of $0.01$. However with
larger mass doubling times the damping rates are still smaller than
those of a cold planet. We explore some very large mass accretion
rates that are not realistic, but might account for very large
luminosities reached after merger events. We find peak values of the
mass accretion at low mass doubling times where the eccentricity and
inclination are boosted most rapidly. The corresponding mass doubling
times are here $4$~kyrs for the eccentricity and $6$~kyrs for the
inclination.

\begin{figure}
  \includegraphics[width=\columnwidth]{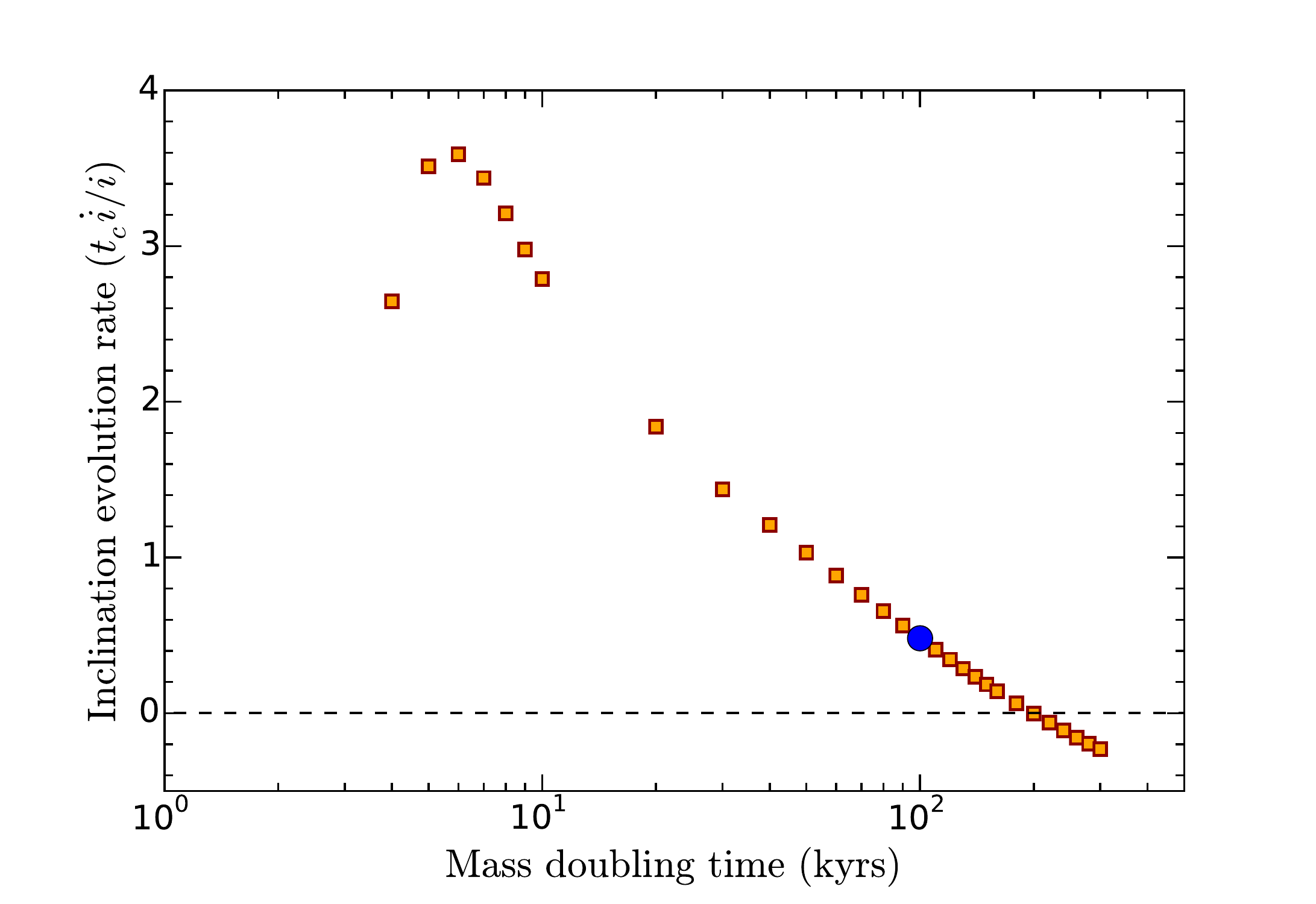}
  \caption{Growth or damping rate of inclination as a function of the mass
    doubling time. The fiducial calculation is indicated with a blue
    disc. For all calculations $i_0=0.01$.}
   \label{fig:inc accretion rate}
\end{figure}

\subsection{Dependence on planetary mass}
\label{sec:planetary mass}
We vary the planetary mass from $5\times10^{-4}$ M$_\oplus$ (about
$4$~percent of the mass of the Moon) to $7.6$ M$_\oplus$, while
keeping the mass doubling time constant. We note, using
Eq.~\eqref{eq:4}, that the luminosity to mass ratio of the planet
scales with $M_\mathrm{p}^{2/3}$ when $\tau$ is kept constant. Our procedure
therefore favours large objects, which have a large luminosity to mass
ratio. Yet we consider that the mass doubling time is a variable
probably more intuitive than the planet's luminosity, and prefer to
explore the planetary mass dependence at fixed $\tau$. The time
derivative of the eccentricity is shown in Figure~\ref{fig:ecc planet
  mass} and that of the inclination in Figure~\ref{fig:inc planet
  mass}. With the fiducial mass doubling time $\tau=10^5$ yrs, the
eccentricity experiences a growth for planetary masses larger than
$M_1=0.25$ M$_\oplus$ and the inclination for planetary masses larger
than $M_1'=0.31$ M$_\oplus$. We also show the behaviour of the
eccentricity for a mass doubling time of $\tau=4\times10^4$~yrs. For
all planetary masses considered, the time derivative of the
eccentricity is larger than for the fiducial value of $\tau$. We find
that, in this case, the eccentricity experiences a growth for
planetary masses larger than $M_2=0.058$~M$_\oplus$. We also find a
drop of the efficiency of both the eccentricity and inclination
driving for planetary masses larger than a few Earth masses, despite
the large luminosity-to-mass ratio implied for these masses by our
constant mass doubling time. For our mass doubling time of
$\tau=10^5$~kyrs, a planet will experience an eccentricity growth
above the value $e_0=0.01$ up to a mass
$M_\mathrm{max}\approx 5.7\;M_\oplus$, and an inclination growth up to
a mass marginally larger than our upper mass limit of $7.6\;M_\oplus$
(extrapolating the data of Fig.~\ref{fig:inc planet mass} we obtain
$M'_\mathrm{max}=8.1\;M_\oplus$).

Note that on the low-mass end of these graphs, the planet is essentially
non-luminous, and we recover the vigorous damping of the eccentricity and
inclination that we found in section~\ref{sec:radiative-disc}, with
dimensionless coefficients larger, in absolute value, than the analytic
estimates of \citet{2004ApJ...602..388T}.

\begin{figure}
  \includegraphics[width=\columnwidth]{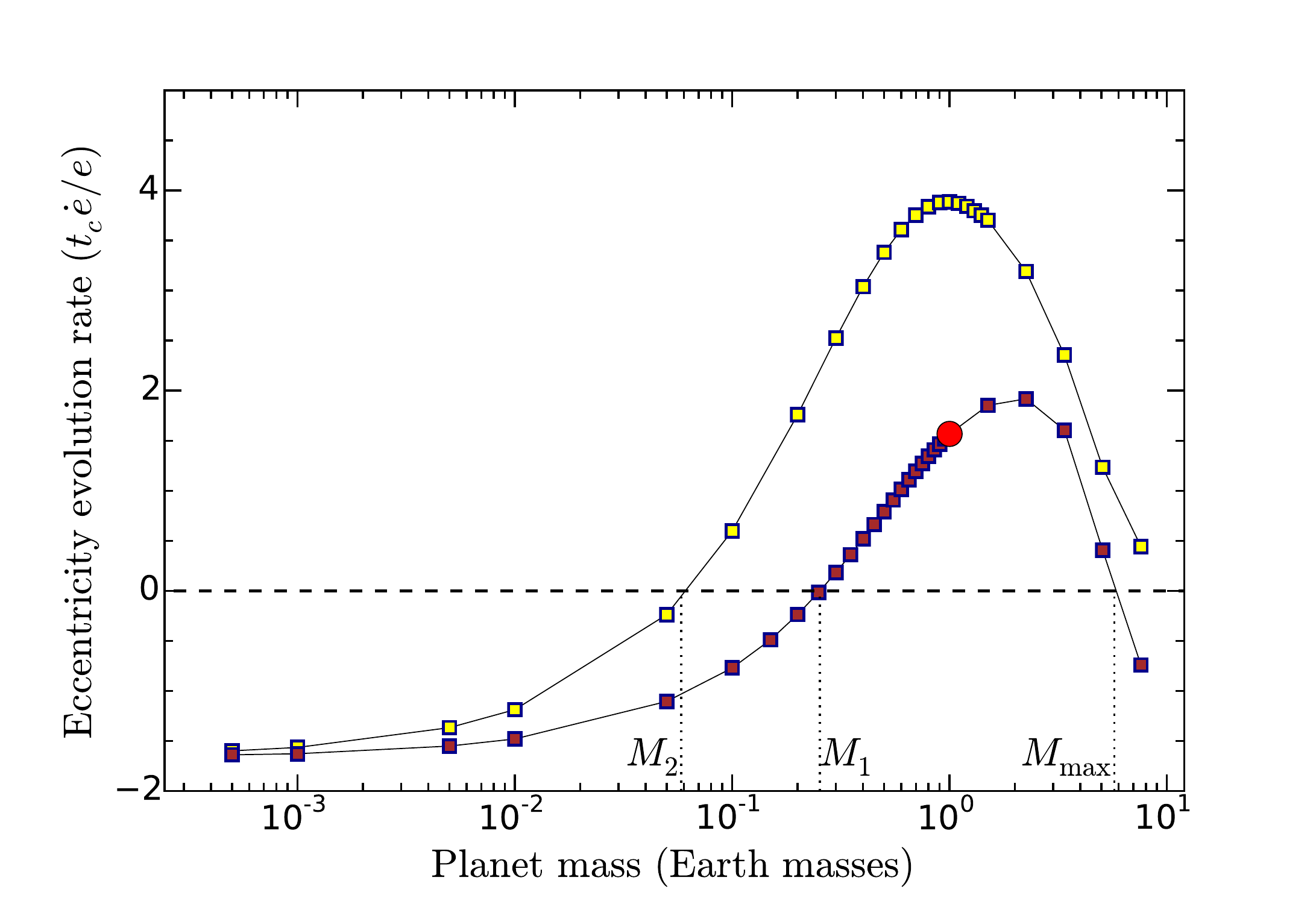}
  \caption{The evolution rate of eccentricity as a function of the planet mass,
    for two different values of the mass doubling time. The purple markers
    correspond to $\tau=10^5$~yrs (the red disc on this series corresponds to
    the fiducial calculation) and the yellow markers correspond to
    $\tau=4\times10^4$~yrs.}
    \label{fig:ecc planet mass}
\end{figure}

\begin{figure}
  \includegraphics[width=\columnwidth]{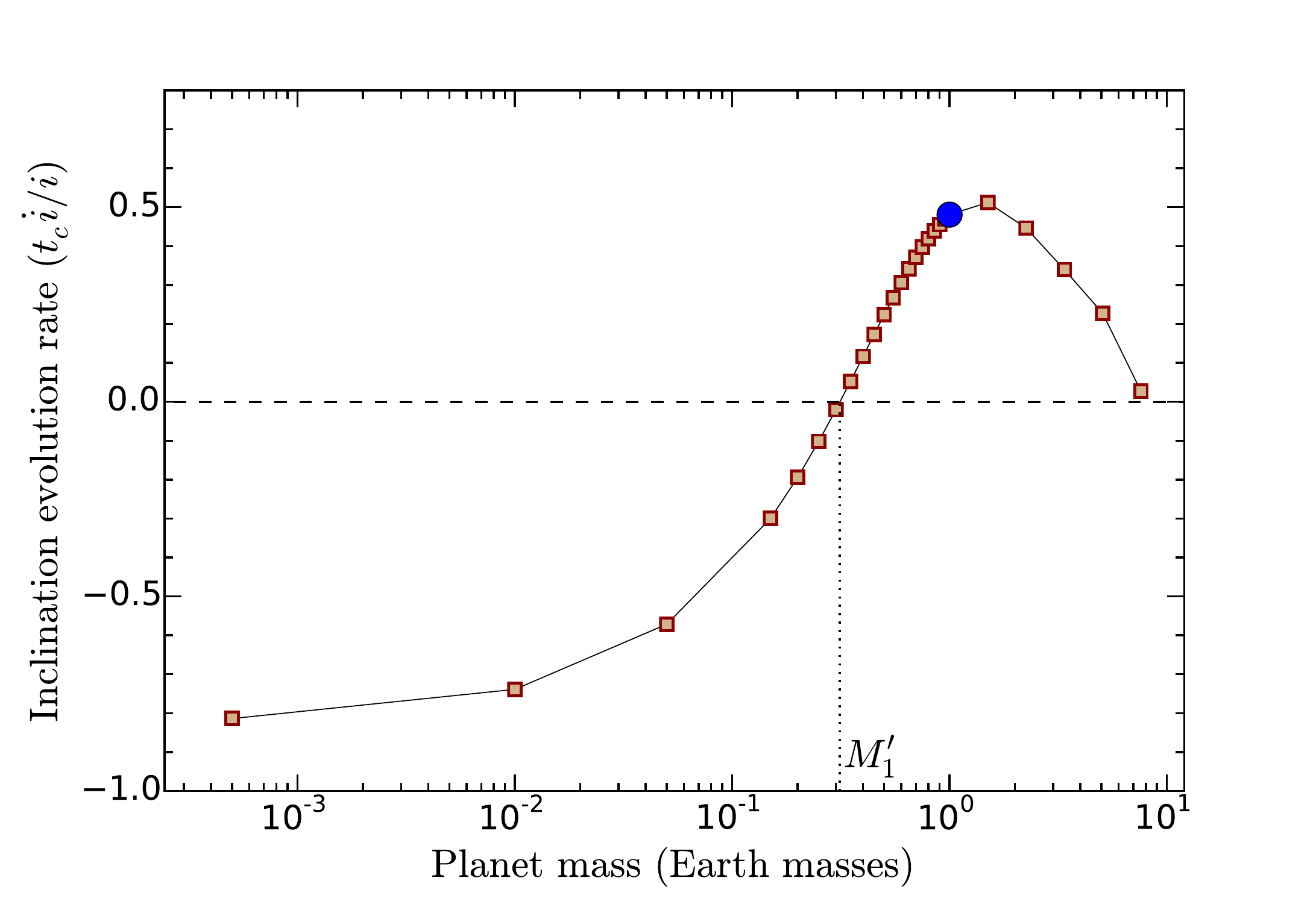}
  \caption{The evolution rate of inclination as a function of the planet mass. 
    The mass doubling time has the fiducial value of $\tau=10^5$~yrs.}
    \label{fig:inc planet mass}
\end{figure}

\subsection{Temperature perturbations in the disc}
\label{sec:temp-pert-disc}
\begin{figure*} 
  \includegraphics[width=\textwidth]{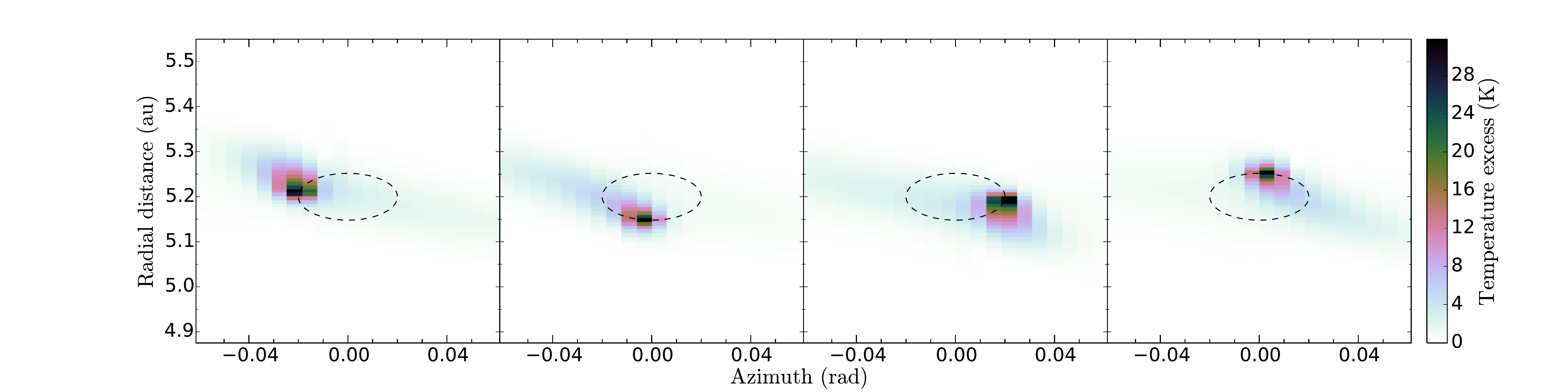}
  \caption{Temperature excess due to the heat release at the disc's
    midplane, for an eccentric planet.  The dotted ellipse represents
    the planet's epicycle. The orbit's guiding centre is at the centre
    of each plot. The second and fourth plots show the planet at the
    pericentre and apocentre, respectively. The time difference
    between successive plots is one fourth of the planet's
    orbital period. A video of this sequence is available at the
    online version of this article.}
   \label{fig:ecc epicycle}
\end{figure*}

We show in Fig.~\ref{fig:ecc epicycle} the temperature difference
between a run with a luminous planet and a run with a non-luminous
planet, at the disc's midplane, when the planet is at four different
locations of its orbital motion. The parameters used here are those of
the fiducial eccentric calculation. In this section only, the planet
is not freely moving in the disc.  Rather, it is held on a fixed
eccentric orbit. Indeed, a luminous and non-luminous planet do not
follow exactly the same trajectory, and subtracting the temperature
fields of two runs where the planet is not exactly at the same
location would be meaningless.

\begin{figure*}
  \includegraphics[width=\textwidth]{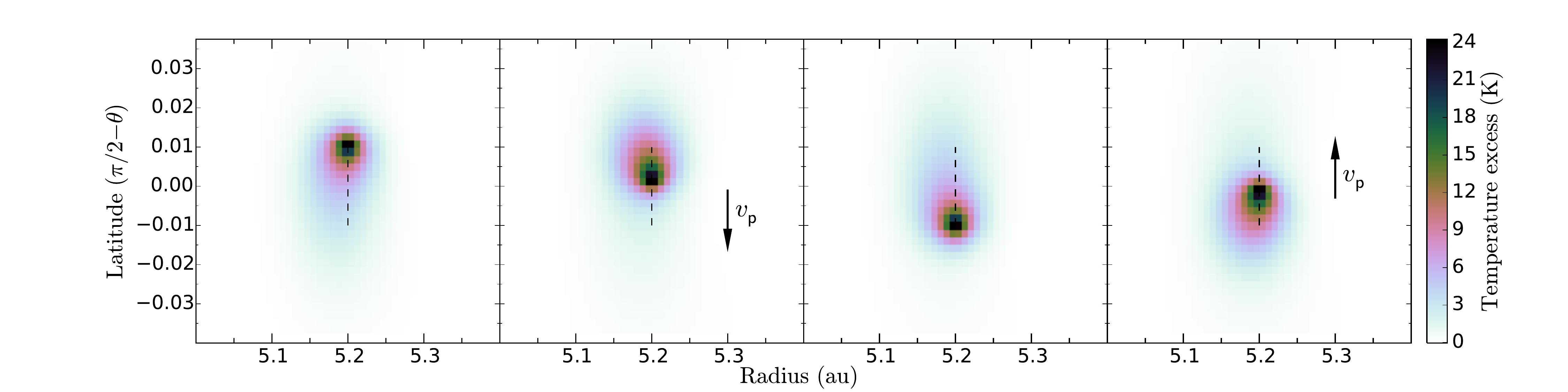}
  \caption{Temperature excess due to the heat release in the
    meridional ($r,\theta$) plane, for an inclined planet.  The dotted
    line represents the planet's trajectory in this plane. The time
    difference between successive plots is one fourth of the planet's
    orbital period.  The arrows show the direction of the planetary
    motion. In the first and third plot, the planet has a vanishing
    vertical velocity.  A video of this sequence is available at the
    online version of this article.}
  \label{fig:inc epicycle}
\end{figure*}

Fig.~\ref{fig:inc epicycle} shows similar plots for the case of the
fiducial calculation of an inclined planet. In both cases, we see a
``hot plume'' trailing the planet. Note that the unperturbed disc's
midplane temperature at the planet's location is $T_0=82$~K, so the
relative temperature perturbation in the plume is at the $O(10^{-1})$
level. In order to compare the size of this plume to the expression given
by \citet{2017MNRAS.465.3175M}, we require an estimate of the disc's
thermal diffusivity. We use the estimate given by \citet[][their
Eq.~16]{2014A&A...564A.135B}:
\begin{equation}
  \label{eq:8}
  \chi=\frac{16\gamma(\gamma-1)\sigma
    T^4}{3\kappa(\rho_0H\Omega_\mathrm{p})^2}\approx 4.6\cdot
  10^{15}\mbox{~cm$^2$~s$^{-1}$ at 5.2~au}
\end{equation}
where $\rho_0$ is the disc's midplane density. According to
\citet{2017MNRAS.465.3175M}, the plume's size depends on the planet's
velocity with respect to the ambient gas. At apo- or pericentre, this
velocity is $a\Omega_\mathrm{p}e/2$.  Inserting this value in Eq.~(24) of
\citet{2017MNRAS.465.3175M}, we get for the plume's cut-off length scale:
\begin{equation}
  \label{eq:9}
  \lambda=\frac{2\chi}{\gamma a\Omega_\mathrm{p}e}\sim 6.5\cdot 10^{-3} a\sim 0.03\mbox{~au}.
\end{equation}
This length only occupies one zone in azimuth, four zones in radius
and three zones in colatitude. The plume is therefore barely resolved,
which suggests that the heating force it exerts on the planet is
underestimated (according to Fig.~1 of \citet{2017MNRAS.465.3175M},
roughly half of the force comes from a distance larger than $2\lambda$
from the planet, and $\sim 80$~\% from a distance larger than
$\lambda$). This cut off scale is much smaller than the disc's
pressure length scale $H$, and also smaller than the planet's
epicyclic excursion $ae$. This is in agreement with the visual
examination of Figs.~\ref{fig:ecc epicycle} and~\ref{fig:inc epicycle}.

The heat released in the planet vicinity takes some time to diffuse
within the plume. This time has been called the response time by
\citet{2017MNRAS.465.3175M}, and has the expression:
\begin{equation}
  \label{eq:21}
  \tau_\mathrm{diff}\sim\frac{\lambda^2}{\chi}\sim\frac{\chi}{\gamma^2V^2},
\end{equation}
where $V$ is the perturber's velocity with respect to the ambient gas.
As the heat diffuses within the plume, the latter is simultaneously
sheared by the Keplerian flow, over a time scale
$(2A)^{-1}=(2/3)\Omega_p^{-1}$, where $A$ is Oort's first
constant. The shear is unimportant if the response time is much
smaller than the shear time scale. In these conditions, the force
exerted by the heated region over the perturber is essentially that
given by a dynamical friction calculation. We refer to this regime as
a headwind-dominated regime. The condition
$\tau_\mathrm{diff} \ll (2/3)\Omega_p^{-1}$ translates into $e\gg e_l$,
with:
\begin{equation}
  \label{eq:22}
  e_l=\frac
  1\gamma\left(\frac{6\chi}{a^2\Omega_p}\right)^{1/2}\approx 0.012,
\end{equation}
where as above we have used $V=ae\Omega_p/2$. For eccentricities below
this limit, the shape of the heated region is strongly affected by the
shear. We refer to this regime as the shear-dominated regime. We
comment that the limit eccentricity $e_l$ can also be worked out by
requiring that the cut-off length $\lambda$ be equal to the distance
between the planet and corotation at peri- or apocentre.

\begin{figure}
  \includegraphics[width=\columnwidth]{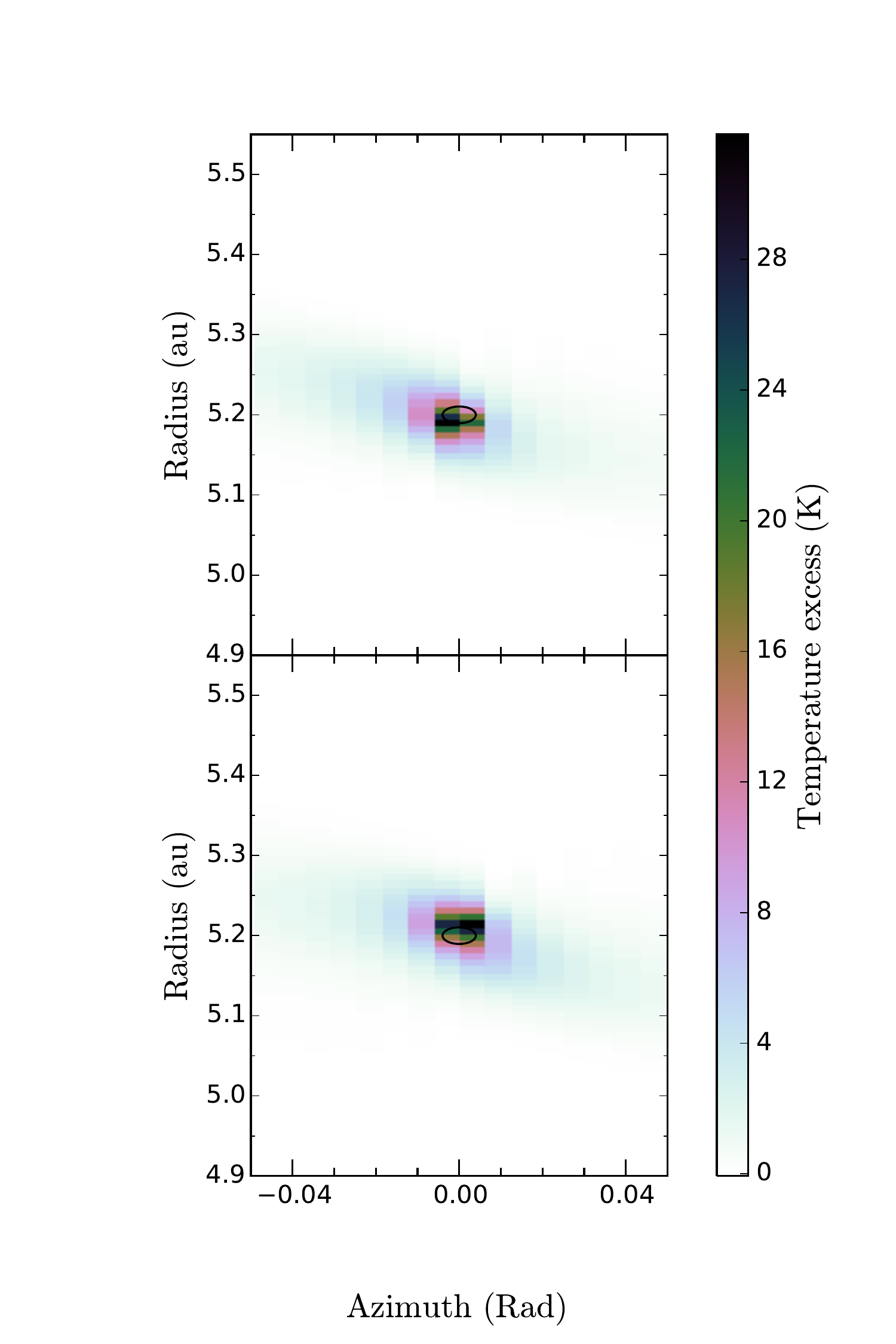}
  \caption{Temperature excess due to the heat release at the disc's 
    midplane, for a low eccentricity planet.  The ellipse represents 
    the planet's epicycle. This figure has been obtained in the same 
    manner as fig.~\ref{fig:ecc epicycle}, except that the planetary 
    eccentricity is here $e=2\cdot 10^{-3}$. The planet is at 
    pericentre in the top plot, and at apocentre in the bottom plot. 
    A video of this case is available at the online version of 
    this article.}
    \label{fig:loweepi}
\end{figure}

In the headwind-dominated regime the plume can essentially be regarded
as a hot trail behind the planet as in Figs.~\ref{fig:ecc epicycle}
and~\ref{fig:inc epicycle}. In the shear-dominated regime, it has a
more complex shape with two lobes sheared apart by the flow: the heat
that has diffused to smaller radii is advected toward positive
azimuth, while the heat that has diffused to larger radii is advected
toward negative azimuth. This is illustrated in
Fig.~\ref{fig:loweepi}. In the limit case of a circular orbit, one
gets two lobes in steady state as found by
\citet{2015Natur.520...63B}. In this case the most pronounced lobe
resides on the same side of corotation as the planet.

\subsection{Long-term behaviour}
\label{sec:long term behavior}
We focus now on the long-term evolution of embedded, hot
planets. Given the large computational cost of each simulation
presented in this section (they represent typically 1-2
months$\cdot$GPU on our cluster), we present a limited number of such
simulations. Note that although the total amount of time spanned by a
given run may amount to a substantial fraction of the mass doubling
time, we keep the planetary mass constant during our simulations. This
inconsistency allows us to separate the behaviour arising from heating
from the behaviour that would arise from an intrinsic mass growth.

\subsubsection{Fiducial parameters} 
\label{sec:fiducial-parameters}
\begin{figure}
  \includegraphics[width=\columnwidth]{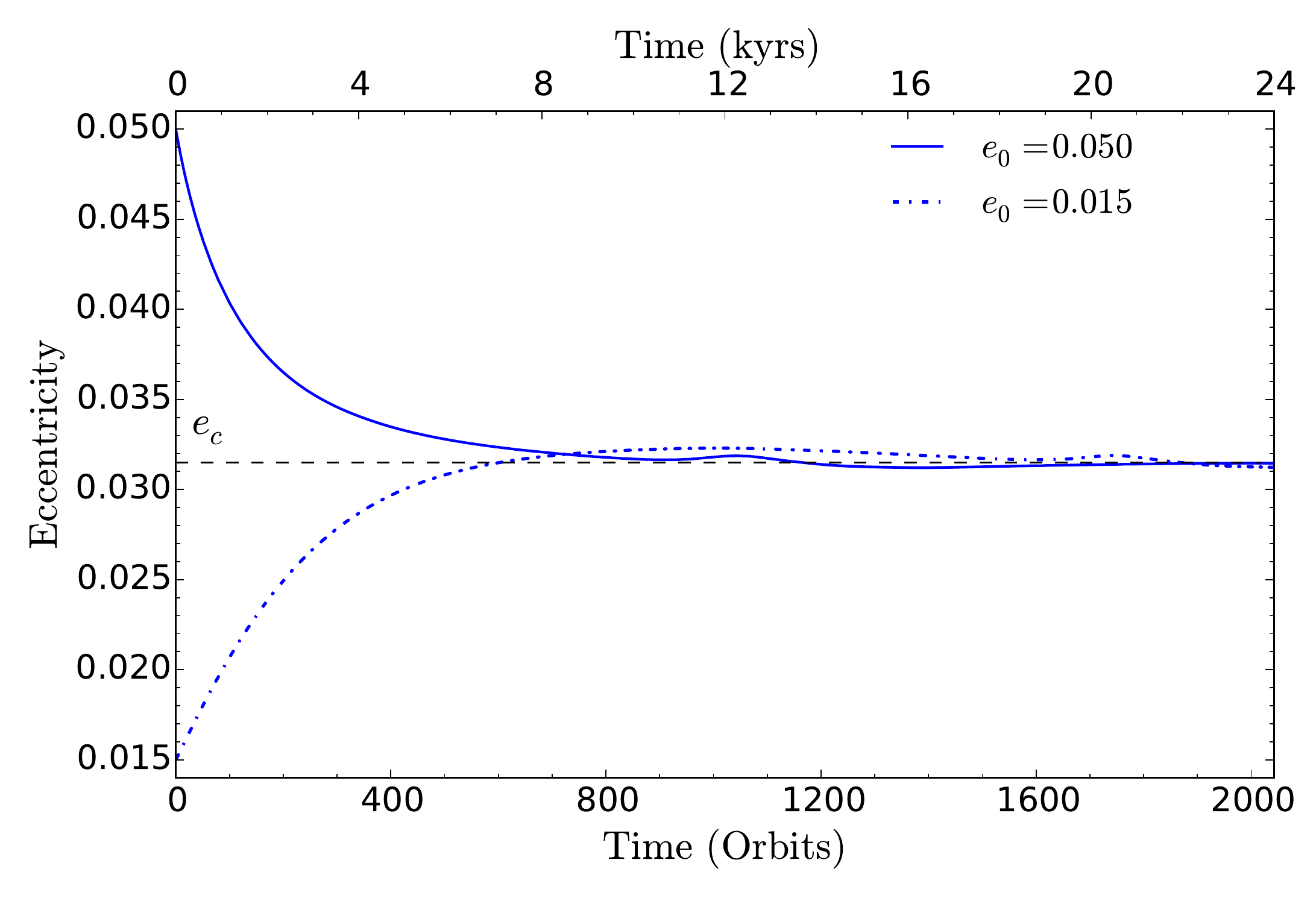}
  \caption{Evolution over 2000 orbits of the eccentricity
    for two different initial values, $e_0=0.015$ and
    $e_0=0.050$. In both cases the eccentricity converges
    towards $e_c$.}
    \label{fig:ecc infinity asympht}
\end{figure}

The study presented in section~\ref{sec:varying initial values}
suggests that the eccentricity and inclination would converge
respectively toward $e_c$ and $i_c$ at larger time, for our fiducial
parameters. We check whether this assumption is
correct. Fig.~\ref{fig:ecc infinity asympht} shows the long-term
behaviour of the eccentricity, which is found to converge indeed
toward $e_c$, regardless of its initial value. Although it does so in
a globally smooth manner, the eccentricity is not strictly monotonous
in time and exhibits low amplitude, non-periodic oscillations over
time.

\begin{figure}
  \includegraphics[width=\columnwidth]{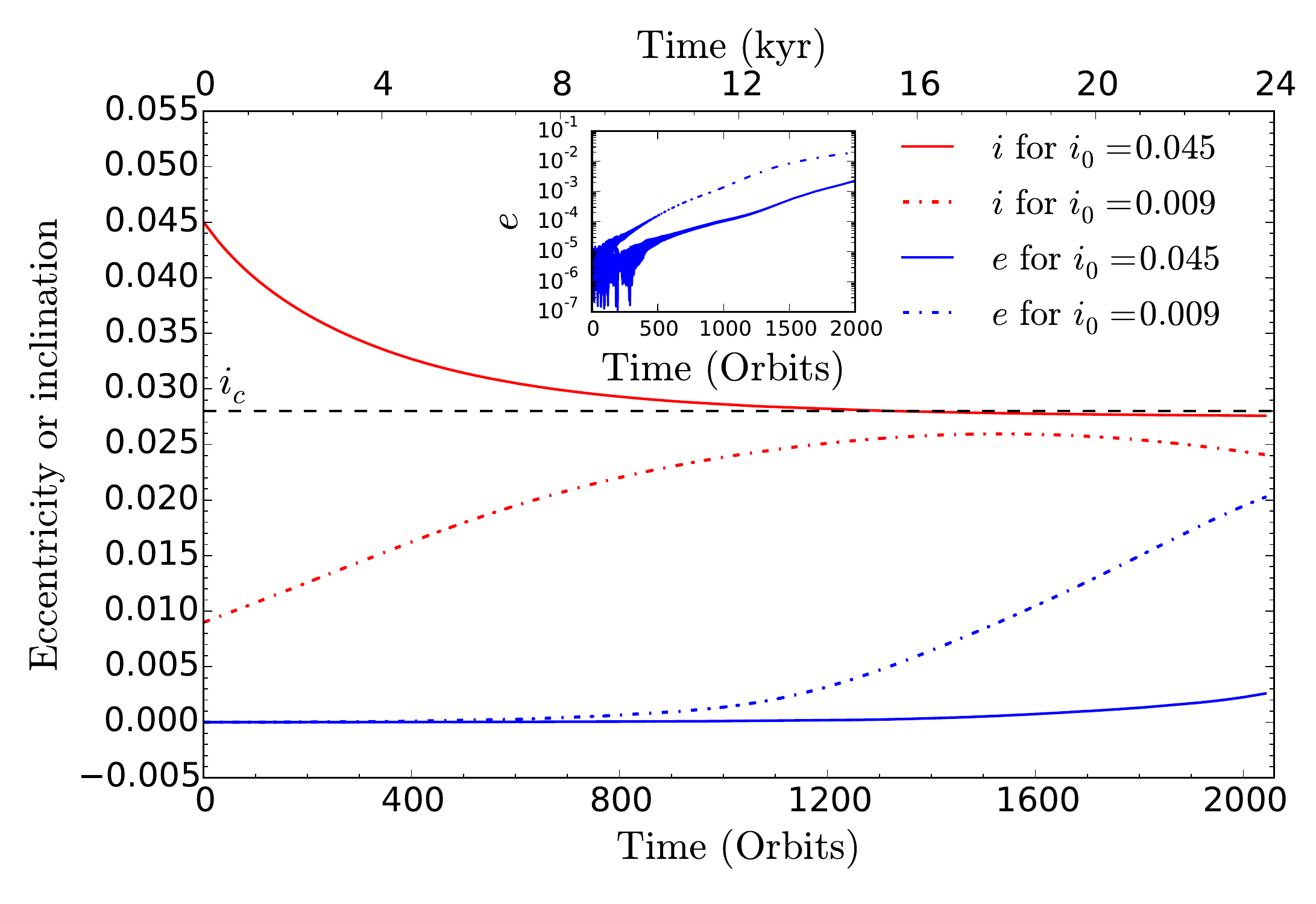}
  \caption{Evolution of the inclination over 2000 orbits, for two
    different initial values: $i_0=0.009$ and $i_0=0.045$. The blue
    lines (initially close to the $x$-axis) show the eccentricities in
    the corresponding runs, with a matching line style. The inset plot
    shows the eccentricity behaviour on a logarithmic scale.}
    \label{fig:inc infinity asympht}
\end{figure}

Fig.~\ref{fig:inc infinity asympht} displays the behaviour of the
inclination for the same disc and planet parameters. Initially, a
trend similar to that of the eccentricity is found, in which the
inclination tends toward the value of $i_c$ found in
section~\ref{sec:varying initial values}. However, the run with a low
initial value of the inclination shows a reversal of the inclination
trend at $t\approx 1500$~orbits. This behaviour occurs when the
planet's eccentricity becomes sizeable. The initial eccentricity of
the planet indeed quickly levels off to a very small, non-vanishing
value given by the perturbations induced in the disc by the planet, of
the order of $10^{-6}$. Subsequently, as can be seen in the inset plot
of Fig.~\ref{fig:inc infinity asympht}, the eccentricity grows
exponentially.
The growth rate of the eccentricity is found to
depend on the inclination for this set of values (a simple explanation
for this could be that the inclination is here comparable to the
disc's aspect ratio, hence the more inclined planet spends more time
in rarefied regions of the disc where the interaction with the latter
is weaker).  The above results suggest that the drivings of the
inclination and eccentricity by the heating are somehow coupled. 

\subsubsection{Further results for a larger luminosity}
\label{sec:furth-results-larg}
We further explore this possibility by comparing the outcome of
simultaneous eccentricity and inclination growth episodes with
different starting values. In order to speed up the time evolution of
the system in our different configurations, we adopt from now on a
mass doubling time of $50$~kyrs, a factor of two shorter than the
fiducial value. All other parameters are those of the fiducial
calculations. A consequence of the planet's larger luminosity is that
the asymptotic values of eccentricity and inclination are larger than
the fiducial ones.
\begin{figure}
  \includegraphics[width=\columnwidth]{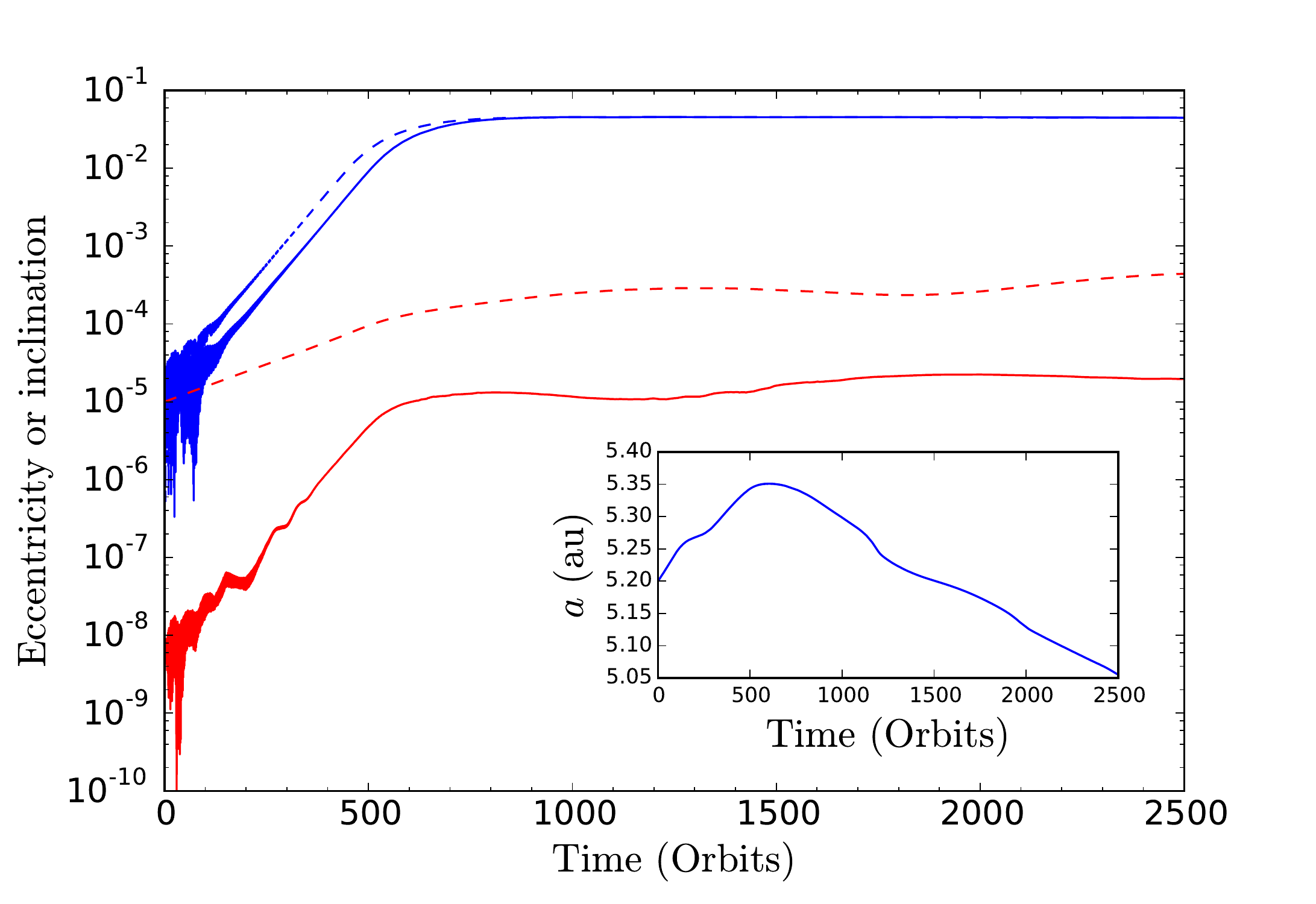}
  \caption{Long-term evolution of the eccentricity and inclination for
    an Earth mass planet with a doubling time $\tau=50$~kyrs in our
    fiducial disc, for two different starting conditions:
    $(e_0,i_0)=(10^{-5},10^{-5})$ (dotted lines) and
    $(e_0,i_0)\sim(10^{-6},10^{-9})$ (solid lines). The two curves in
    the upper part of the diagram show the eccentricity (in blue in
    the electronic version), while the other two curves (in red in the
    electronic version) show the inclination with a matching line
    style. The inset plot shows the time behaviour of the semi-major
    axis of the planet of the second case.}
    \label{fig:ltb2}
\end{figure} 
We show in Fig.~\ref{fig:ltb2} the time evolution of the eccentricity
and inclination for two different starting conditions: one in which
$e_0=i_0=10^{-5}$, and another one in which these values were set to
zero, but quickly level off to very small values because of the
perturbation induced by the planet for the eccentricity (around
$\sim 10^{-6}$) or because of numerical noise for the inclination
(around $10^{-9}$). In both cases the eccentricity displays a phase of
exponential growth over several decades in the shear-dominated regime
(which we had anticipated from the examination of Fig.~\ref{fig:ecc
  variation}), and then levels off toward an asymptotic value of
$0.044$. The inclination, on the contrary, exhibits a very different
behaviour: after a phase of exponential growth (with a different
growth rate in each case), it levels off at very low values, orders of
magnitude smaller than its asymptotic value which can be inferred from
an analysis similar to that of section~\ref{sec:varying initial
  values}, and which is $\sim 0.04$. It is noteworthy that the
inclination levels off when the eccentricity reaches sizeable values
($\gtrsim 0.01$). This suggests that a value of the eccentricity close
to the nominal value quenches the inclination growth. 

Fig.~\ref{fig:ltb2} also shows the time behaviour of the semi-major
axis of the planet which has initially
$(e_0,i_0)\sim(10^{-6},10^{-9})$. In agreement with the results of
\citet{2015Natur.520...63B}, the planet initially migrates outwards,
and reverses its migration once its eccentricity has reached a
sizeable value of $\sim 0.01$. Although we have not investigated the
reasons for this behaviour, it appears compatible with a substantial
fraction of the total torque arising from a positive corotation
torque, which is quenched once the radial excursion $ae$ is larger
than the half-width of the horseshoe region
$x_s\approx 8\cdot 10^{-3}a$ \citep{2012MNRAS.419.2737H}. We comment
that this migration behaviour may be affected by the bias mentioned in
section~\ref{sec:reduct-azim-extent}, and that the total variation of
the semi-major axis over the whole extent of this long-term run is
minute, as anticipated.

\begin{figure}
  \includegraphics[width=\columnwidth]{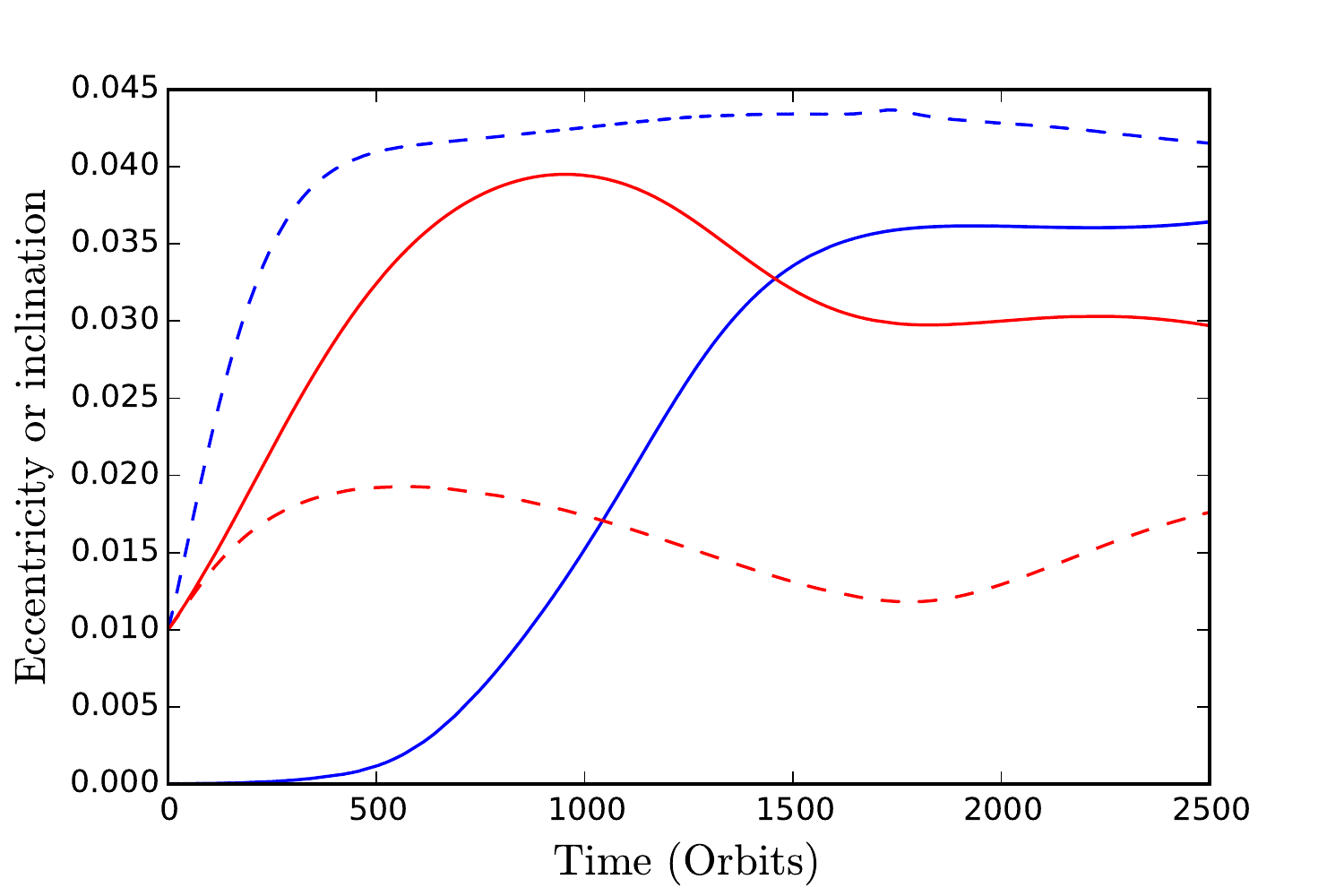}
  \caption{Results for calculations similar to those of
    Fig.~\ref{fig:ltb2} for $(e_0,i_0)=(0.01,0.01)$ (dashed lines) and $(e_0,i_0)=(\sim 10^{-6},0.01)$ (solid
    lines). The eccentricity (inclination) curves appear in blue (red) on the electronic version.}
    \label{fig:ltb3}
\end{figure}

We conclude this
section by showing results in Fig.~\ref{fig:ltb3} for similar
calculations, this time with $(e_0,i_0)=(10^{-6},0.01)$ and with
$(e_0,i_0)=(0.01,0.01)$. The former shows a behaviour similar to the
one seen in Fig. \ref{fig:inc infinity asympht}, with an initial
growth of the inclination followed by a mild decay once the
eccentricity has reached a sizeable value. At larger time the
eccentricity and inclination appear to have plateaued toward nearly
constant values, both smaller than their asymptotic values taken
separately, the eccentricity being marginally larger than the
inclination. The other calculation presented in this figure, which
starts with $e_0=i_0=0.01$, shows that the eccentricity grows toward a
nearly constant value between $0.04$ and $0.045$, whereas the
inclination has an initial growth rate smaller than in the other
calculation, and never exceed $0.02$. This clearly illustrates the
interdependence of the time evolution of the eccentricity and
inclination.

\subsection{A toy model}
\label{sec:toy-model}
We present here a toy model which reproduces some of the
characteristics of the eccentricity and inclination growth in the
headwind dominated regime. Under these conditions, the force on the
planet arising from the heat release can be obtained by a dynamical
friction calculation \citep{2017MNRAS.465.3175M}.

The Hamiltonian of the planet in a frame corotating with its guiding
centre is given, in spherical coordinates, by:
\begin{eqnarray}
  \label{eq:10}
  H&=&\frac 12M_\mathrm{p}\dot r^2+\frac 12M_\mathrm{p}r^2\sin^2\theta\dot\phi^2+\frac
       12M_\mathrm{p}r^2\dot\theta^2\nonumber\\
&&-\frac 12M_\mathrm{p}r^2\sin^2\theta\Omega_\mathrm{p}^2-\frac{GM_\star M_\mathrm{p}}{r},
\end{eqnarray}
where $\phi$ is the azimuthal angle between the guiding centre and the
planet. We assume from now on that the angular
velocity of the guiding centre is fixed, or equivalently that the
planet does not migrate. The migration timescale is typically two to
three orders of magnitude larger than the eccentricity and inclination
evolution timescales \citep{arty93b,tanaka2002,2004ApJ...602..388T},
even when the planet releases heat \citep{2015Natur.520...63B}, so
this assumption is reasonable. When the planet is subjected to a
non-conservative force $(F_r,F_\phi,F_\theta)$, its Hamiltonian varies in
time at the rate:
\begin{equation}
  \label{eq:11}
  \dot H=\dot r F_r+r\sin\theta\dot\phi F_\phi+r\dot\theta F_\theta,
\end{equation}
where the right hand side represents the work of the force in the
frame corotating with the guiding centre. The planetary orbit, to
first order in $e$ and $i$, reads:
\begin{equation}
  \label{eq:12}
\left\{
\begin{aligned}
  r&=&a[1+e\cos(\Omega_\mathrm{p}t+\varphi)]\\
  \phi&=&2e\sin(\Omega_\mathrm{p}t+\varphi)\\
  \theta&=&\frac{\pi}{2}-i\sin(\Omega_\mathrm{p} t),\\
\end{aligned}
\right.
\end{equation}
where $\varphi$ is an arbitrary phase and where we have adopted,
without loss of generality, $t=0$ on the ascending line of nodes. Upon
substitution of these expressions in Eq.~\eqref{eq:10}, we obtain:
\begin{equation}
  \label{eq:13}
H=H_c+\frac 12a^2\Omega_\mathrm{p}^2(e^2+i^2),
\end{equation}
where $H_c=-(3/2)a^2\Omega_\mathrm{p}^2$ is the value of~$H$ when the planet is
on a circular, non-inclined orbit.  The impact of the heating force on
the orbital elements can be evaluated by averaging its work in the
rotating frame over one orbital period.  \citet{2017MNRAS.465.3175M}
have shown that in the low Mach number limit the force does not depend
on the perturber's velocity. Here we assume the force to be a
constant $F_0$, and to be directed along the velocity vector of the
planet relative to the ambient gas. When the planet is inclined and
non-eccentric, the work is obtained readily as:
\begin{equation}
  \label{eq:14}
  W_i=4F_0ai.
\end{equation}
Equating this work to the variation of $H$ over one orbital period, we
can write the time derivative of the inclination:
\begin{equation}
  \label{eq:15}
  \frac{di}{dt}=\frac{2F_0}{\pi aM_\mathrm{p}\Omega_\mathrm{p}}\approx 0.64\frac{F_0}{aM_\mathrm{p}\Omega_\mathrm{p}}
\end{equation}
We now consider an eccentric, non-inclined planet. Using the notation:
$x=r-a$ and $y=a\phi$, we can write to lowest order in $e$, with an
arbitrary choice of the time origin:
\begin{equation}
  \label{eq:16}
  \mathbfit{q}=(x,y)=[ae\sin(\Omega_\mathrm{p}t),2ae\cos(\Omega_\mathrm{p}t)]
\end{equation}
The gas velocity is $[0,(-3/2)\Omega_\mathrm{p}x]$, hence the planet's velocity
in the gas frame reads:
\begin{equation}
  \label{eq:17}
  \mathbfit{v'}=(v_x',v_y')=ae\Omega_\mathrm{p}[\cos(\Omega_\mathrm{p}t),-(1/2)\sin(\Omega_\mathrm{p}t)].
\end{equation}
The work of the heating force over the epicycle is therefore:
\begin{eqnarray}
  W_e=\oint F_0\frac{\mathbfit{v'}\cdot d\mathbfit{q}}{v'}&=&\int_0^{\frac{2\pi}{\Omega_\mathrm{p}}}\frac{F_0ae\Omega_\mathrm{p}
                                                                         dt}{[\cos^2(\Omega_\mathrm{p}t)+\sin^2(\Omega_\mathrm{p}
                                                                         t)/4]}\nonumber\\
\label{eq:18}
&=&8.63F_0ae.
\end{eqnarray}
From this work we can deduce the time derivative of the
eccentricity:
\begin{equation}
  \label{eq:19}
  \frac{de}{dt}\approx 1.37\frac{F_0}{aM_\mathrm{p}\Omega_\mathrm{p}}. 
\end{equation}
Eqs.~\eqref{eq:15} and~\eqref{eq:19} show that in the
headwind-dominated regime, the eccentricity and inclination should
experience a linear growth. This is compatible with the graphs of
Fig.~\ref{fig:ltb3}.

We could extent this study by adding to the work of the
non-conservative force the contribution of the disc's tide, so as to
match the damping rates given by \citet{2004ApJ...602..388T} or those,
more recent, of \citet{2007A&A...473..329C}. Our preliminary study of
section~\ref{sec:radiative-disc} shows nevertheless that the damping
rates in a radiative disc are at odds with those given for isothermal
discs. On general grounds, however, we note that the exponential or
nearly exponential damping of $e$ ($i$) by the disc's tide implies
that the work of the tide should be negative and scale as $e^2$
($i^2$), i.e. faster than that of the heating force. For values of $e$
($i$) lower than $e_c$ ($i_c$), the work of the heating force
dominates and the orbital element grows, until reaching the value
($e_c$ or $i_c$) at which the tide's work cancels out the work of the
heating force. At this point the orbital element remains constant in
time.

The work of Eq.~\eqref{eq:18} is only $11$~\% smaller than the
estimate that would be obtained assuming that the force is tangent
everywhere to the epicycle. This remark allows to understand why the
impact of the force is larger on eccentricity than on inclination: it
is simply because the epicycle circumference is significantly larger
than (twice\footnote{Over one orbital period the motion is
  alternatively downward and upward.}) the length of the arc described
by an inclined planet, for configurations where the eccentricity and
inclination have same values. This results in the heating force exerting
more work on an eccentric planet, over one orbital period.

In this toy model the ratio of the eccentricity to inclination time
derivative is $2.2$, smaller than the ratio of $\sim 3$ that we found
in our simulations. Many factors can account for this difference:
\begin{itemize}
\item Our toy model calculation assumes that the response time of the
  heating force is vanishingly small.
\item It also assumes the heating force to be independent of
  altitude, whereas the sound speed and thermal diffusivity (variables
  that come into play in the expression of the heating force) depend
  on altitude.
\item We have a low, anisotropic resolution, so the heating force in
  our numerical simulations can have a different value for different
  directions of motion.
\item Our toy model neglects the damping effect of the disc tide
  on the eccentricity and inclination.
\end{itemize}
Nevertheless, our calculation suggests that the stronger impact on
eccentricity should be a rather general outcome, and not a particular
result arising from the specifics of our fiducial disc model.

If we now consider a planet that is both eccentric and inclined, it is
still subjected to a heating force of magnitude $F_0$, the horizontal
projection of which drives the eccentricity, while its vertical
projection drives the inclination. The effect on both orbital elements
is therefore necessarily weaker than the effect obtained when
considering each orbital element in isolation, \emph{i.e.} when the
planet is either only eccentric or only inclined. Furthermore, the
ratio of the vertical to horizontal component of the heating force is
$\sim i/e$. If the planet has an eccentricity much larger than its
inclination, it is left with a very small vertical component of the
force to drive the inclination, which is counteracted at low values by
the disc's tide.  This simple model therefore broadly accounts for the
salient features of the long-term behaviours that have been presented in
sections~\ref{sec:fiducial-parameters}
and~\ref{sec:furth-results-larg}. It also suggests that the asymptotic
value of the orbital elements should depend on $L/M_\mathrm{p}$ if they result
from a balance between the heating force (which scales as $LM_\mathrm{p}$) and
the disc's tide (which scales as $M_\mathrm{p}^2$).

\subsubsection{Planetary mass}
\label{sec:planetary-mass}
In order to check this hypothesis, we study here the asymptotic value
of the eccentricity and inclination as a function of the planetary
mass, for our fiducial parameters. This value is determined by a
dichotomic search of the value of $e$ or $i$ that leads to a
negligible variation of the orbital element over~$5$ orbits. We have
seen in section~\ref{sec:long term behavior} that this is not
necessarily the value toward which the orbital element will converge,
especially for the inclination, which, in a calculation with small
initial values of the orbital elements, is overrun by the eccentricity
and levels off at low values. It provides nonetheless an estimate of
the maximum value that one can expect for the orbital elements.

\begin{figure}
  \includegraphics[width=\columnwidth]{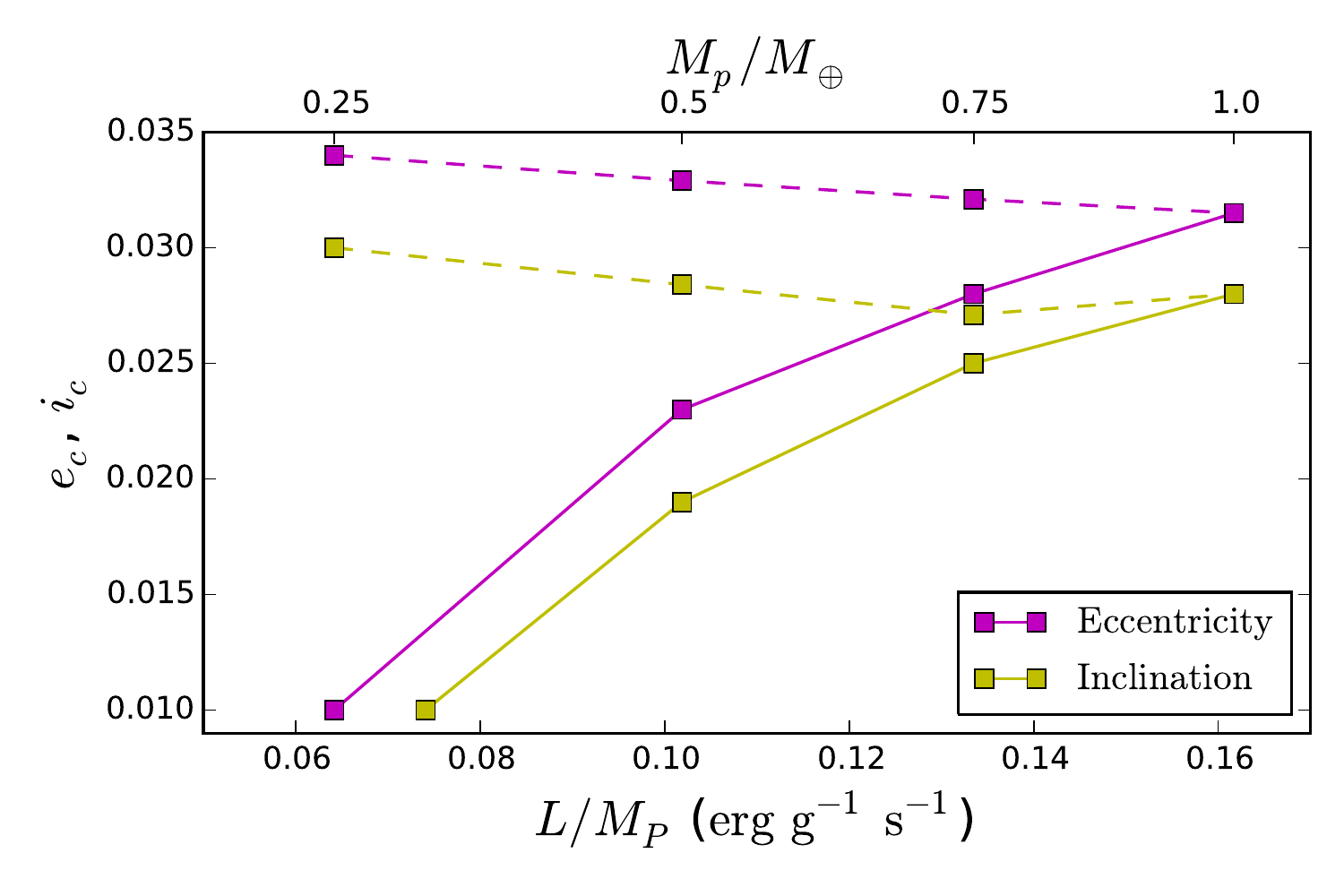}
  \caption{Asymptotic values of the eccentricity (magenta in the
    electronic version) and inclination (yellow in the electronic
    version) as a function of the planetary mass for a constant mass
    doubling time (solid lines) and for a constant luminosity to mass
    ratio (dashed lines). The bottom axis only refers to the solid
    lines, since the data of the dashed lines has been obtained for a
    constant value of $L/M_\mathrm{p}=0.16$~erg~g$^{-1}$~s$^{-1}$.}
  \label{fig:ecic}
\end{figure}

We show two kinds of dependencies in Fig.~\ref{fig:ecic}. One is
obtained by varying the planetary mass with a constant mass doubling
time. The other one is obtained by varying the planetary mass and the
mass doubling time so as to have a constant luminosity to mass ratio,
equal to that of the fiducial run
($L/M_\mathrm{p}=0.16$~erg.g$^{-1}$.s$^{-1}$).  This second series
shows nearly constant orbital parameters, which implies that, for
given disc parameters, their asymptotic value essentially depends on
the luminosity to mass ratio, as suggested in
section~\ref{sec:toy-model}. On the contrary, in the first series, the
dependence of the orbital parameters on the planetary mass reflects
the variation of the luminosity to mass ratio of the planet. We
comment that our fiducial luminosity to mass ratio is nearly two
orders of magnitude larger than the specific heating rate due to the
decay of $^{26}$Al at the CAI formation \citep[e.g.][and
refs. therein]{2006mess.book..555G}. Therefore, an embryo solely
heated by $^{26}$Al decay would not be luminous enough to experience a
growth of eccentricity or inclination.

\section{Discussion}
\label{sec:discussion}
As we have mentioned in section~\ref{sec:temp-pert-disc}, the
resolution employed in this study is barely sufficient to describe the
planetary hot trail. Fig.~1 of \citet{2017MNRAS.465.3175M} shows that
an accurate estimate of the heating force requires to resolve scales
that are much smaller than the cut-off length~$\lambda$, which is not
the case here. Using Eqs.~\eqref{eq:15} and~\eqref{eq:19} with the
estimate of the heating force given at low Mach number by
\citet{2017MNRAS.465.3175M} leads to growth rates several times in
excess of those found in section~\ref{sec:fiducial calculation}, which
suggests that the heating force is indeed largely underestimated in
our present setup.

On the other hand, a more refined treatment of the planetary accretion
may imply a reduced the planetary luminosity and ultimately lower
eccentricity and inclination growth rates.  Our simple expression of
the luminosity (Eq.~\ref{eq:2}) assumes that all the potential energy
of the accreted material is radiated instantaneously, and translates
into an effective temperature:
\begin{equation}
  \label{eq:20}
  T_\mathrm{eff}=1220\left(\frac{M_\mathrm{p}}{M_\oplus}\right)^{1/4}\left(\frac{\rho_\mathrm{p}}{3\,\mathrm{g~cm}^{-3}}\right)^{1/4}\left(\frac{\tau}{100\,\mathrm{kyrs}}\right)^{-1/4}\,\mathrm{K}.
\end{equation}
In a different context, that of the atmosphere of Earth-like planets
after giant impacts, \citet{2014ApJ...784...27L} show that the
effective temperature of the planet may be substantially smaller than
its surface temperature due to the blanketing by the atmosphere,
delaying the planet's cooling. Should an accreting embryo be subjected
to a similar blanketing, the efficiency of our driving mechanism would
be lowered. For a given maximal effective temperature, smaller embryos
would have a larger luminosity to mass ratio, since the luminosity
scales as $R_\mathrm{p}^2$ and the mass as $R_\mathrm{p}^3$.

The Bondi sphere of our fiducial $1\;M_\oplus$ planet is unresolved.
The flow is non-linear and has a complex structure at the sub-Bondi
scale \citep{2016arXiv161009375F,2015arXiv150503152F}. The force
arising from the hot gas within the Bondi sphere is not captured by
our analysis and is presently unknown. This underlines the need for
very high resolution calculations in the planet vicinity, which would
probably need either nested meshes \citep{2016MNRAS.460.2853S} or
freely moving meshes \citep{2014MNRAS.445.3475M} to meet the demanding
resolution requirements. We note that if the diffusion timescale of
the heat across the Bondi radius is smaller than the acoustic time
across the Bondi radius, the dynamical impact of the heat release
within the Bondi radius should be limited. This occurs when
$M_\mathrm{p} < 4\chi c_s/G=2.5\;M_\oplus$ for our fiducial disc.

We find a cut-off of the excitation of eccentricity and inclination
past values of $\gtrsim 1\;M_\oplus$. This cut-off is similar to that
found by \citet{2015Natur.520...63B} for migration. From
considerations on the yield of the conversion of the planet's
luminosity into kinetic energy, \citet{2017MNRAS.465.3175M} argue that
the magnitude of the heating force must decay when
$GM/\chi c_s \gtrsim 1$. This threshold is in broad agreement with the
numerical value quoted above. A deeper understanding of this cut-off
definitely requires a resolved calculation of the flow at the sub-Bondi
scale.

\citet{2017MNRAS.465.3175M} have speculated that planets with a large
luminosity to mass ratio could undergo indefinite eccentricity
driving, resulting in the planet being supersonic with respect to the
gas. We have not found such outcome for any reasonable value of the
planetary luminosity in our setup. However, the plume size decreases
when the velocity of the planet relative to the gas increases. Our
lack of supersonic planets may therefore simply reflect our lack of
resolution, and stresses again the need for calculations at very high
resolution.

\section{Conclusions}
\label{sec:conclusions}
We find that Earth-sized planetary embryos heated by accretion and
embedded in opaque protoplanetary discs can experience eccentricity
and inclination growth to values that are comparable to the disc's
aspect ratio, over timescales of the order of a few kyrs, for
luminosities corresponding to mass doubling times of order of
$10^5$~yrs.  The asymptotic value depends on the luminosity to mass
ratio of the planet. However, a more refined treatment of the
planetary luminosity, and higher resolution calculations are required
to accurately determine the ultimate values of the eccentricity and
inclination.

The origin of this growth is the appearance of a hot, underdense
region in the planet vicinity. The size of this region, for the
parameters that we considered, is significantly smaller than the
disc's pressure scale length. As a consequence, the transition between
a shear-dominated regime (in which this region is sheared apart by the
Keplerian flow) to a headwind-dominated regime (for which the
Keplerian shear becomes unimportant) occurs for values of the
eccentricity significantly smaller than the disc's aspect ratio (at
$e\sim 0.01$ for our set of parameters). In the headwind-dominated
regime, we expect a heating force similar to that described by
\citet{2017MNRAS.465.3175M} to act on the planet.  The value reached
at larger time by the eccentricity or inclination is set by the
balance between the heating force, which excites the orbital element,
and the disc's tide, which damps it.

We find that the drivings of the eccentricity and inclination by
heating are coupled at larger time. Our interpretation is as
follows. In the headwind-dominated regime, the heating force is nearly
constant. Its horizontal projection drives the eccentricity, while its
vertical projection drives the inclination. The values reached at
larger time by these orbital elements, when the planet is both
eccentric and inclined, are therefore smaller than the values reached
by a planet that is either only eccentric, or only inclined, since
only a fraction of the (constant) heating force contributes to the
driving of each of these orbital element. This accounts for the
coupling of the eccentricity and inclination evolutions.

We find that the eccentricity has a growth rate approximately three
times larger than that of the inclination. When a planet has initially
a very small eccentricity and inclination, the eccentricity overruns
the inclination, and the planet reaches the headwind-dominated regime
with a very low value of the inclination. It is then subjected to a
nearly horizontal heating force, which quenches any further growth of
the inclination. The outcome is therefore a planet with a significant
eccentricity (comparable to the disc's aspect ratio) and a very small
inclination. When, on the contrary, the planet has initially a
sizeable inclination, it will remain sizeable, and at larger time the
eccentricity and inclination will have same order of magnitude.

Extensions of this work could be an assessment of the actual
luminosity of embryos, and the derivation of analytical prescriptions
for the effect of heat release on eccentricity and inclination, so as
to evaluate its impact on scenarios of dynamical relaxation and
collisions of a set of protoplanetary embryos.

As a side result, we find damping rates of eccentricity and
inclination on a non-luminous planet, in a radiative disc, that are
significantly larger than those found in isothermal discs.

\section*{Acknowledgements}
H. Eklund wishes to thank the Linnaeus-Palme exchange program for
their support, and the \emph{Instituto de Ciencias F\'\i sicas} of
UNAM for support and hospitality. The authors acknowledge CONACyT grant 178377 and
UNAM's PAPIIT grant 101616. The authors thank the referee for a constructive report
and J. Szul\'agyi for a thorough reading of a first draft of this manuscript.

%%%%%%%%%%%%%%%%%%%%%%%%%%%%%%%%%%%%%%%%%%%%%%%%%%



\bibliographystyle{mnras}

%%% Alternatively you could enter them by hand, like this:
%%% This method is tedious and prone to error if you have lots of
%%% references

% Don't change these lines
\bsp	% typesetting comment
\label{lastpage}
\end{document}